\documentclass[useAMS,usenatbib,final]{mn2e}
\usepackage{graphics}
\usepackage{epsfig}
\usepackage{color}
\usepackage{xspace}
\usepackage{times}

\newcommand{\Ha}{\ifmmode {\rm H}\alpha \else H$\alpha$\fi\xspace}
\newcommand{\Hb}{\ifmmode {\rm H}\beta \else H$\beta$\fi\xspace}
\newcommand{\Hg}{\ifmmode {\rm H}\gamma \else H$\gamma$\fi\xspace}
\newcommand{\Hd}{\ifmmode {\rm H}\delta \else H$\delta$\fi\xspace}

\newcommand{\Hii}{\ifmmode \rm{H}\,\textsc{ii} \else H~{\sc ii}\fi}

\newcommand{\nii}{\ifmmode [\rm{N}\,\textsc{ii}] \else [N~{\sc ii}]\fi}

\newcommand{\oi}{\ifmmode [\rm{O}\,\textsc{i}] \else [O~{\sc i}]\fi}

\newcommand{\neiii}{\ifmmode [\rm{Ne}\,\textsc{iii}] \else [Ne~{\sc iii}]\fi}
\newcommand{\hei}{\ifmmode [\rm{He}\,\textsc{i}] \else [He~{\sc i}]\fi}
\newcommand{\oii}{\ifmmode [\rm{O}\,\textsc{ii}] \else [O~{\sc ii}]\fi}

\newcommand{\oiii}{\ifmmode [\rm{O}\,\textsc{iii}] \else [O~{\sc iii}]\fi}

\newcommand{\sii}{\ifmmode [\rm{S}\,\textsc{ii}] \else [S~{\sc ii}]\fi}
\newcommand{\siii}{\ifmmode [\rm{S}\,\textsc{iii}] \else [S~{\sc iii}]\fi}

\newcommand{\tl}{$\langle\log t_\star\rangle_L$}
\newcommand{\tm}{$\langle\log t_\star\rangle_M$}

\newcommand{\zl}{$\log \langle Z_\star\rangle_L$}
\newcommand{\zm}{$\log \langle Z_\star\rangle_M$}

\title[Nature via nurture]
         {Semi-empirical analysis of Sloan Digital Sky Survey galaxies: 
         IV.~A~nature via nurture scenario for galaxy evolution}

\author[Mateus et al.]
{Ab{{\'\i}}lio Mateus$^{1}$\thanks{E-mail: abilio@astro.iag.usp.br},
Laerte Sodr\'e Jr.$^{1}$, 
Roberto Cid Fernandes$^{2}$,
Gra\.zyna Stasi\'nska$^{3}$\\
$^{1}$Departamento de Astronomia, IAG-USP, Rua do Mat\~ao 1226, 05508-090, S\~ao Paulo, Brazil\\
$^{2}$Depto.\ de F\'{\i}sica - CFM - Universidade Federal de Santa Catarina, Florian\'opolis, SC, Brazil\\
$^{3}$LUTH, Observatoire de Meudon, 92195 Meudon Cedex, France}

\begin{document}

\pagerange{\pageref{firstpage}--\pageref{lastpage}} \volume{000} \pubyear{2006}

\maketitle

\begin{abstract}
We investigate the environmental dependence of stellar population properties of galaxies in the local universe. Physical quantities related to the stellar content of galaxies are derived from a spectral synthesis method applied to a volume-limited sample containing more than 60 thousand galaxies ($0.04 < z < 0.075$; $M_r \le -19.9$), extracted from the Data Release~4 of the Sloan Digital Sky Survey (SDSS). Mean stellar ages, mean stellar metallicities and stellar masses are obtained from this method and used to characterise the stellar populations of galaxies. The environment is defined by the projected local galaxy density estimated from a nearest neighbour approach. We recover the star formation--density relation in terms of the mean light-weighted stellar age, which is strongly correlated with star formation parameters derived from \Ha. We find that the age--density relation is distinct when we divide galaxies according to luminosity or stellar mass. The relation is remarkable for galaxies in all bins of luminosity. On the other hand, only for an intermediate stellar mass interval (associated to a transition in galaxy properties) the relation shows a change in galaxy properties with environment. Such distinct behaviours are associated to the large stellar masses of galaxies with the same luminosity in high-density environments. In addition, the well known star formation--density relation results from the prevalence of massive systems in high-density environments, independently of galaxy luminosity, with the additional observed downsizing in galaxy formation, in which the star formation is shifted from massive galaxies at early times to low-mass galaxies as the universe evolves. Finally, our results support that a natural path for galaxy evolution proceeds \emph{via} a nurture way, in the sense that galaxy evolution is accelerated in denser environments, that took place mainly at high-redshifts.
\end{abstract}
\label{firstpage}

\section{Introduction}

Galaxy environment has a fundamental role in the evolutionary paths followed by galaxies.
It is well known that galaxy populations change according to the environment in which they are found. The most classical evidence of such environmental dependence is the morphology-density relation \citep[e.g.][]{dressler80,whitmore93}. The fraction of galaxies with distinct morphological types (essentially spirals, lenticulars and ellipticals) correlates strongly with the local galaxy density, with high-density environments being populated preponderantly by early-type galaxies. This is related to the dependence of the fraction of star-forming galaxies with the environment, based in the presence or not of emission lines, such as \Ha\ and \oii, in galaxy spectra \citep[e.g.][]{carter01,hashimoto98,mateus04}. Such dependence is also intrinsically related to the reduced gaseous content of galaxies in denser regions \citep{solanes96,bravo00,goto03}.

Recent studies have found that the star formation rate (SFR) of galaxies is the most sensitive parameter to galaxy environment, declining strongly in high-density regions of galaxy clusters \citep[e.g.][]{lewis02,gomez03,rines05}, whereas structural parameters are less affected by environment \citep{kauffmann04}. The relation environment-SFR has also an additional dependence on galaxy luminosity, being steeper for fainter galaxies \citep{tanaka04}. However, when one restricts the analysis to star-forming galaxies the median SFR of this class of objects seems to be unaffected by environment, although the fraction of such galaxies decreases with increasing local density \citep{balogh04}.

The most plausible way to account for these trends is to assume that galaxy properties (mainly those related to star formation and gas properties) are affected by environment through well known physical mechanisms acting on galaxies. This path, linked directly to the environment, gives origin to a \textit{nurture} perspective for galaxy evolution.
In fact, several physical mechanisms were already proposed and studied to account for the evolutionary trends discussed above. Interactions between the intragalactic
and intergalactic medium, including gas removal and evaporation
\citep[\textit{ram pressure stripping};][]{gunn72,fujita99,vollmer01}, and the suppression of the accretion of gas-rich materials in the neighbourhood of the galaxy \citep[\textit{starvation};][]{LTC,bekki01a}, are typical mechanisms claimed to affect star formation properties of galaxies infalling onto denser regions.

Another path by which galaxy evolution proceeds is related to the initial conditions established during the formation epoch of galaxies, which could, in principle, account for the relations between galaxy properties and environment. This path gives origin to a
\textit{nature} perspective driving galaxy evolution. As expected from a biased galaxy formation scenario \citep[e.g.][]{cen93}, massive galaxies are formed earlier preferentially in high-density regions; in opposition, low mass galaxies would be formed later with a smooth distribution in the density field. Thus, in this scenario there is, naturally, an expected relation between age of formation (or mass) and environmental density. In fact, hierarchical galaxy formation models successfuly reproduce the morphology-density relation \citep[e.g.][]{benson01}, demonstrating that it is present even if the physical mechanisms mentioned above are not considered.

Additionally, studies based on high-redshift galaxies \citep[e.g. ][and others]{cowie96,kodama04,juneau05} and also on local galaxies (Kauffmann et al. 2003a; Heavens et al. 2004; Mateus et al. 2006, hereafter SEAGal II) have revealed that the existence of a `downsizing' in galaxy formation is extremely important in analysis involving the star formation properties of galaxies. These results suggest that
massive galaxies have stopped to form stars at earlier times, with low mass systems comprising a large fraction among galaxies with ongoing star formation. These trends are also recovered by recent high-resolution simulations \citep[e.g.][]{weinberg04} and semi-analytic models for galaxy formation \citep[e.g.][]{menci05,delucia05}. However, the physical origin of the downsizing is still a subject of debate \citep[e.g.][]{delucia05, scannapieco05, bundy06}.

Here we will shed some light into the discussion concerning galaxy formation and evolution by investigating the role of environment on the stellar population properties of galaxies in the nearby universe. Other works have increased our understanding on this issue through distinct approaches \citep[e.g.][]{kauffmann04,thomas05,poggianti05,clemens06}. In this paper, we will use the results of the application of a spectral synthesis method to a volume limited sample of SDSS galaxies. This method is able to recover important physical properties of galaxies from their spectra, including mean stellar ages, mean stellar metallicities and stellar masses, among others. The details of such approach were discussed by Cid Fernandes et al. (2005) (hereafter SEAGal I) and its results have been recently used in SEAGal II, where we have studied the bimodality of the galaxy population.

In this work we show that the relations between galaxy properties and environment, well characterised by the age--density relation, are distinct when we divide galaxies according to luminosity or stellar mass. This is a result of the different mass-to-light ($M_\star/L$) ratios of galaxies in distinct environments. In dense regions galaxies of same luminosity tend to have higher $M_\star/L$ ratios. We argue that a natural path emerges from these results but via a nurture way in the sense that massive galaxies are formed preferentially in high-density regions (where the evolution was accelerated) in the high-$z$ universe (when the star formation rate was high), since their stellar populations also tend to be the oldest ones and they inhabit the densest regions we observe in the local universe.

The paper is organised as follows. Section 2 describes the data used in this
work, the galaxy sample definition and a brief overview of the spectral
synthesis method used in our analysis. Section 3 describes some parameters
related to star formation activity in galaxies that will be discussed in this work, and Section 4 defines the environmental parameter, namely the projected local galaxy density. In Section 5 we investigate the environmental dependence of the stellar population properties of galaxies. Finally, in Section 6 we discuss the implications of our findings in the context of galaxy evolution, and in Section 7 we summarise the main results of this work.

\section{The data}

The data used in this work were extracted from the Sloan Digital Sky
Survey available publicly in its Data Release 4 \citep[DR4;][]{DR4}.
The SDSS is an ambitious project planned to study the large scale
structure of the universe as well as other relevant subjects of
extragalactic astronomy. The survey aims to cover one-quarter of the
entire sky (mainly in the Northern Hemisphere) obtaining photometric
information in five optical bands ($u$, $g$, $r$, $i$, and $z$) for
millions of objects and spectroscopic data for about $10^6$ galaxies
brighter than $r = 17.77$.

In this section we describe the volume-limited sample of SDSS galaxies
used in our analysis and the application of a spectral synthesis method
to the spectroscopic data aiming to obtain some fundamental physical
properties of galaxies in the local universe. In this work we use the
following values of the cosmological parameters:
$H_0=70$ km s$^{-1}$ Mpc$^{-1}$, $\Omega_M=0.3$ and $\Omega_\Lambda=0.7$.

\subsection{Galaxy sample}

We built a volume limited sample of galaxies from the SDSS DR4. This sample follows the same criteria used by \citet{strauss02} to define the Main Galaxy Sample, consisting of galaxies with $r$-band Petrosian magnitudes $r \le 17.77$ and $r$-band Petrosian half-light surface brightnesses $\mu_{50} \le 24.5$~mag~arcsec$^2$. We have selected
galaxies with redshifts in the range $0.04 < z < 0.075$ with $r$-band extinction-corrected
absolute magnitudes $M_r \le -19.9$, corresponding roughly to $M^*_r + 1.5$. The absolute magnitudes adopted here are $k$-corrected by using the {\sevensize KCORRECT v3\_2} code provided by \citet{blanton03kcorrect}. The lower redshift limit was chosen here to avoid the undesirable presence of aperture effects in our analysis which are related to the use of small 3\arcsec\ fibres in the SDSS to obtain the spectroscopic data \citep{kewley05}.
We have detected a minor fraction of galaxies with multiple spectral data, from which we have considered only those with best signal-to-noise ratio ($S/N$) in the $g$-band. These selection criteria result in a volume limited sample containing 63659 galaxies.

\begin{figure*}
\resizebox{\textwidth}{!}{\includegraphics{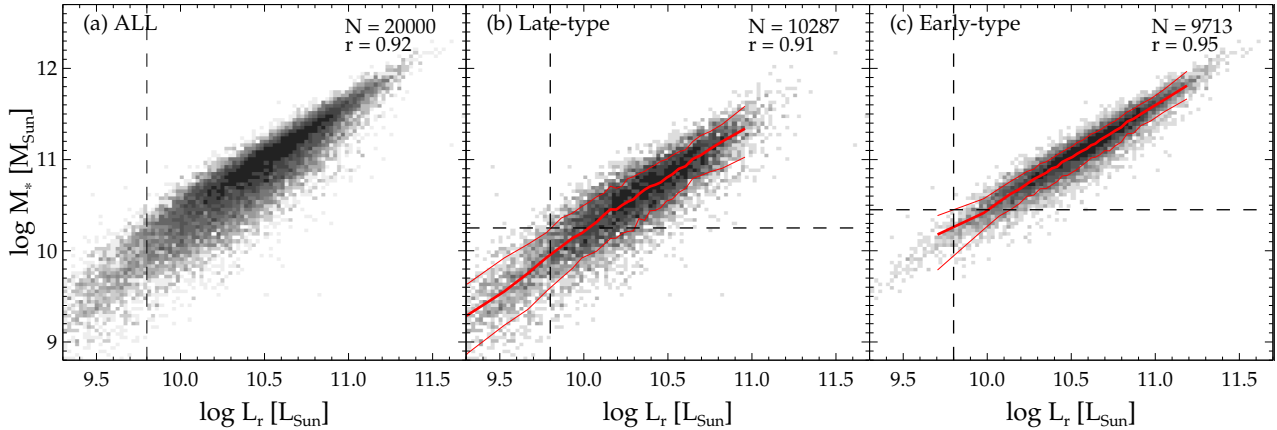}}
\caption{Mass-luminosity relation for (a) all galaxies in the 20k
sample, and for (b) late-type galaxies with $D_n(4000) < 1.67$ and (c)
early-type galaxies with $D_n(4000) > 1.67$. Solid lines are the median values (thick lines) and 10-90 percentiles (thin lines) for the relations. The luminosity limit for our sample ($\log L_r/L_\odot = 9.80$) is shown as vertical dashed lines, while the mass values above which our sample will be considered complete are shown as horizontal dashed lines. The numbers at the top-right are the count of galaxies in each panel and the Spearman correlation coefficient.}
\label{Fig:mass_luminosity_20k}
\end{figure*}

\subsection{Overview of the spectral synthesis method}\label{sec:synthesis}

We used the {\sevensize STARLIGHT} code (described in detail in
\citet{seagal1} and updated in \citet{mateus06}) to obtain physical
parameters of galaxies directly from their spectra (essentially from
continuum and absorption lines).  {\sevensize STARLIGHT} fits an
observed spectrum with a combination of $N_\star$ Simple Stellar
Populations (SSP) from the evolutionary synthesis models of
\citet{bc03}. Here, we used a base containing $N_\star = 150$
elements, spanning 6 metallicities -- $Z=0.005, 0.02, 0.2, 0.4, 1$ and
2.5 $Z_\odot$ -- and 25 ages, between 1 Myr and 18 Gyr.  Stellar
extinction is modelled as due to foreground dust, with the extinction
law of \citet*{cardelli89} with $R_V=3.1$, and parametrised by the
V-band extinction $A_V$. Line of sight stellar motions are modelled by
a Gaussian distribution centred at velocity $v_\star$ and with
dispersion $\sigma_\star$. Regions around emission lines and bad
pixels are excluded from the analysis. \citet{mateus06} describes
numerical and other technical aspects of the code.

We apply our synthesis method to the galaxy sample described in the
last section. The SDSS spectra cover a wavelength range of 3800--9200
\AA, have mean spectral resolution $\lambda/\Delta\lambda \sim 1800$,
and were taken with 3 arcsec diameter fibres.  Prior to the spectral
fits, the spectra are corrected for Galactic extinction with the maps
given by \citet*{schlegel98} and the extinction law of
\citet{cardelli89}, shifted to the rest-frame and resampled from 3400
to 8900 \AA~in steps of 1~\AA.

The result of the procedure is a list of parameters for each galaxy.
The most important for the discussion in this paper are:
a) the dust-corrected stellar mass inside the fibre; the total stellar
mass ($M_\star$) is computed \textit{a posteriori} after correcting
for the fraction of luminosity outside the fibre assuming that the 
galaxy mass-to-light ratio does not depend of the radius;
b) the mean stellar age, weighted either by light, \tl, or by mass
\tm; c) the mean stellar metallicity, light and mass-weighted, \zl\ and
\zm, respectively; d)  the V-band stellar extinction $A_V^\star$; and e)
the velocity dispersion, $\sigma_\star$. Besides, by subtracting the synthesised
spectrum from the observed one, we get a ``pure-emission'' residual
spectrum, which is useful for analysis of the galaxy emission lines.

\begin{figure*}
\centerline{\includegraphics[scale=0.7]{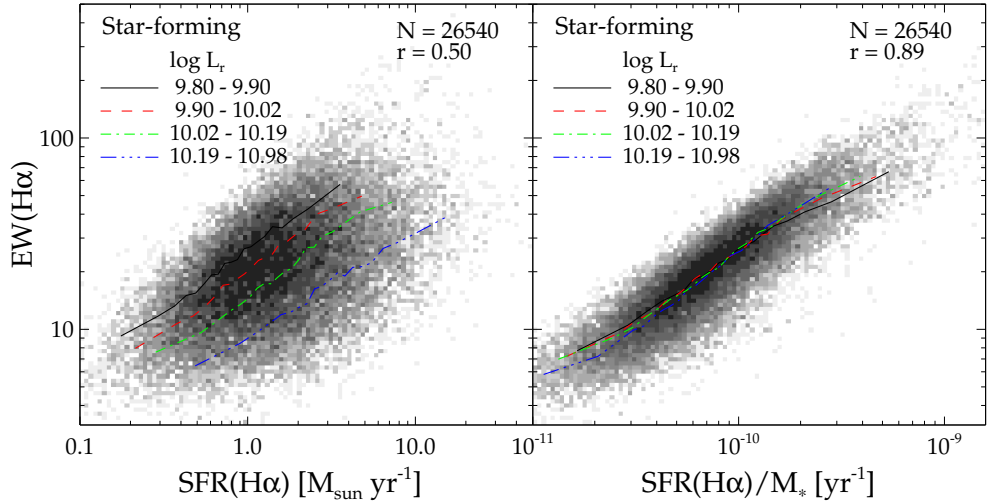}}
\caption{Relations among the tracers of star formation activity based on the \Ha\ emission line. Median values of EW(\Ha) as a function of SFR(\Ha) (left panel) and
SSFR(\Ha) (right panel) are shown in bins of $r$-band galaxy luminosity chosen to contain the same number of objects. In this and in the following figures, the median values for each luminosity bin are shown as different lines. The legend in the top-left corner shows the range of each bin in which the median values have been evaluated. The numbers at the top-right are the count of galaxies in each panel and the Spearman correlation coefficient.}
\label{fig:EWHa_versus_SFR_Ha}
\end{figure*}
\begin{figure*}
\centering\includegraphics[scale=0.7]{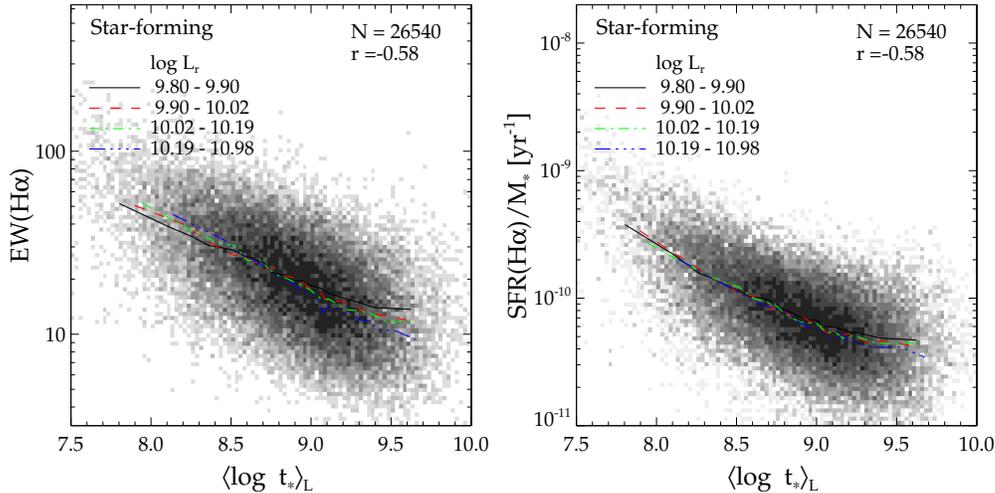}
\caption{Median values of EW(\Ha) and SSFR(\Ha) as a function of
light-weighted mean stellar age in bins of galaxy luminosity. The numbers at the top-right are the count of galaxies in each panel and the Spearman correlation coefficient.}
\label{fig:EWHa_versus_age}
\end{figure*}

\subsection{Completeness issues}\label{sec:completeness}

The volume-limited sample adopted in this work secures a high completeness
for galaxies brighter than $M_r \le -19.9$. However, in this work we are also interested in the properties of galaxies divided according to stellar mass, thus a high completeness in this parameter is needed in order to avoid any bias originated from the sample selection procedure.

We investigate this issue with the help of the mass-luminosity relation of
galaxies from a flux-limited sample containing 20000 objects (hereafter
referred as `20k sample') selected at random. The total stellar masses are
computed after correcting the stellar masses obtained from the spectral
synthesis by aperture effects. As for the main sample, this is done by dividing
the stellar masses by $(1-f)$, where $f$ is the fraction of the total galaxy
luminosity in a given band outside the fibre. Here this correction is made by
using total and fibre Petrosian magnitudes in the $z$-band.

Fig.~\ref{Fig:mass_luminosity_20k} shows the mass-luminosity relation for all
galaxies in the 20k sample, and for early and late-type galaxies
distinguished according to the 4000~\AA\ break index measured following
the definition given by \citet{balogh99} and referred as $D_n(4000)$. In SEAGal II we have shown that the optimal value to separate early and late galaxy types is $D_n(4000) = 1.67$. In this figure we also show the median values (solid lines) and the 10--90 percentiles (dashed lines) of the stellar mass in bins of galaxy luminosity.

In order to select a subsample of galaxies with high completeness in stellar mass we need to compute the value of $M_\star$ above which it will be complete. Adopting the galaxy luminosity limit for our sample ($M_r \le -19.9$, corresponding to $\log L_r/L_\odot \ge 9.8$, shown in Fig.~\ref{Fig:mass_luminosity_20k} as vertical dashed lines), and considering the upper 90 percentile of the distributions for each galaxy type shown in Fig.~\ref{Fig:mass_luminosity_20k}, we found that our galaxy sample will be nearly complete for late-type galaxies with $M_\star \ge 1.8 \times 10^{10}$ M$_\odot$, and for early-type galaxies with $M_\star \ge 2.8 \times 10^{10}$ M$_\odot$. By using these limits (shown as horizontal dashed lines in Fig.~\ref{Fig:mass_luminosity_20k}), we then selected a subsample containing 49453 objects, with a high completeness in stellar mass. Thus, when studying relations involving $M_\star$, we will adopt this subsample in order to avoid any bias introduced by our sample selection procedure. It is convenient to summarise here the samples that will be used in the rest of the paper. In  Table~\ref{table:samples} we show the basic statistics of the volume-limited and stellar mass-limited samples, and also of a subsample of star-forming galaxies that will be analysed in the next section.

\begin{table}
\centering
\begin{tabular}{lccccc}
\hline \hline
Sample  & Number & Per cent \\
\hline
Volume-limited		& 63659 & 100.0  \\
$M_\star$-limited	& 49453 & 77.7 \\
Star-forming galaxies  	& 26540 & 41.7 \\
\hline \hline
\end{tabular}
\caption{Summary of the samples adopted in this work.}
\label{table:samples}
\end{table}

\section{The star formation activity in galaxies}\label{sec:star_formation}

In this section we describe some usual tracers of star formation activity in
galaxies based on the \Ha\ emission line, and how they are
related to spectral synthesis products. Our main intention here is to define these
quantities and show that the light-weighted mean stellar age behaves as a good tracer of
current star formation in galaxies.

\subsection{Emission line related products}

The equivalent width (EW) of the \Ha\ emission line is an useful parameter
when studying the star formation properties of galaxies. This quantity is
related to the ratio between the amount of star formation occurring in a
galaxy in the last $\sim 10^7$ yr and its past integrated star formation
history. Since the measurement of EW(\Ha) does not require flux-calibrated spectra nor
intrinsic reddening correction, it has been widely used as an indicator of
recent star formation activity in local galaxies, for instance in works related to the
environmental dependence of star formation properties of galaxies
\citep[e.g.][]{gomez03,lewis02,rines05}.

A more direct and useful tracer of star formation rate in galaxies is the
\Ha luminosity \citep[e.g.][]{kennicutt98}. When
appropriately corrected for underlying absorption and dust extinction, and in
the case of fibre-based spectra also for aperture effects, the \Ha SFR
estimate is found to be consistent with radio, far-infrared, and $u$-band
estimates \citep{hopkins03}. In our case, the underlying stellar absorption is
automatically subtracted when computing the residual spectra
(as discussed in Section~\ref{sec:synthesis}).
The \Ha SFR, corrected by aperture effects, is then simply derived by using the
expression adapted from \citet{kennicutt98} for a Chabrier IMF (0.1-100 M$_\odot$)
and with the prescriptions given by \citet{hopkins03}:
\begin{equation}
{\rm SFR}_{\Ha}(M_\odot \, {\rm yr}^{-1}) = 5.22 \times 10^{-42} \,
L(\Ha) \, 10^{-0.4(r - r_{\rm fibre})},
\end{equation}
where $L(\Ha)$ is the observed luminosity of \Ha (in ergs s$^{-1}$) corrected
by nebular extinction with the \Ha/\Hb intrinsic intensity ratio, $r$ is the
$r$-band  Petrosian magnitude representing the total galaxy flux, and
$r_{\rm fibre}$ is the $r$-band fibre magnitude.\footnote{The latter term in
this equation is used to correct the SFR estimate by aperture effects by
assuming that the \Ha luminosity inside the fibre is characteristic of the
whole galaxy, and that the continuum flux at the \Ha wavelength is well
represented by the flux at the effective wavelength of the $r$-band filter
\citep[see details in][]{hopkins03}.}
Note that since we have corrected the \Ha\ luminosities for extinction with the help of
the \Ha/\Hb ratio, only galaxies with these two lines measured with significant $S/N$
ratio will have their SFR(\Ha) computed. Consequently, only galaxies with higher levels of star formation activity will have their star formation rates evaluated.

Another interesting quantity related to star formation in
galaxies is the SFR per unit stellar mass, or specific star formation rate (SSFR(\Ha)),
which is associated to the strength of the current burst of star formation
relative to the underlying stellar mass of a galaxy \citep[e.g.][]{guzman97}.
In our case, this parameter is easily obtained by dividing the
SFR(\Ha) (in M$_\odot$ yr$^{-1}$) by the dust-corrected stellar mass,
$M_\star$.

In order to illustrate the quantities described above, in 
Fig.~\ref{fig:EWHa_versus_SFR_Ha} we show the relation between the EW(\Ha)
and the SFR(\Ha), and that between the EW(\Ha) and the specific \Ha\ SFR.
This figure also shows the median values of each relation in
bins of $r$-band galaxy luminosity. It is important to stress here that we only consider galaxies with \Ha\ measured in emission with a $S/N$ greater than 3 (following \citealt{seagal1}). In addition, we also restrict this analysis to galaxies
with \Ha\ emission coming from normal star-forming regions by excluding hosts of active
galactic nuclei in the same way as done by \citet{mateus06}. In
Fig.~\ref{fig:EWHa_versus_SFR_Ha} we note a clear dependence on galaxy luminosity for the
relation between EW(\Ha) and SFR(\Ha), in the sense that for a given value of EW(\Ha),
brighter galaxies have higher SFRs. This luminosity dependence disappears when
dealing with the specific SFR.

\subsection{Light-weighted mean stellar age as an indicator of star formation}

The mean light-weighted stellar age of a galaxy is related to the formation
epoch of massive and bright stars, frequently associated to starbursts.
Hence, low values of \tl\ reflect the recent activity of intense star
formation operating in a given galaxy, whereas high values of it indicate
that most of the galaxy light comes from older stellar populations,
with less significant, or even none, recent star formation episodes. As
this quantity is obtained by our spectral synthesis approach in a robust way
we will inspect its ability in tracing star formation activity of
galaxies.

In Fig.~\ref{fig:EWHa_versus_age} we show the relation between \tl\ and the
parameters discussed previously: EW(\Ha) and SSFR(\Ha). We do not show the relation between \tl\ and SFR(\Ha) since this latter parameter has a dependence on galaxy luminosity, related to the amount of stellar mass recently formed in a galaxy.
As in Fig.~\ref{fig:EWHa_versus_SFR_Ha}, these relations are also shown in
bins of galaxy luminosity (i.e., it is an \emph{extensive} quantity, whereas \tl, EW(\Ha) and SSFR(\Ha) are all \emph{intensive} quantities). We clearly note that this mean stellar age is very sensitive to star formation activity estimated via \Ha\ products. The relations with EW(\Ha) and SSFR(\Ha) show high Spearman correlation coefficients ($-0.58$ for both relations). Hence, the \tl\ figures out as a good tracer of recent star formation activity in galaxies, and since we have measured it for all galaxies in our sample (unlike H$\alpha$ based indicators) it is well suited for analysis involving the whole galaxy population, even for galaxies without emission lines.

\section{The environment}

\begin{figure}
\centerline{\includegraphics[width=\columnwidth]{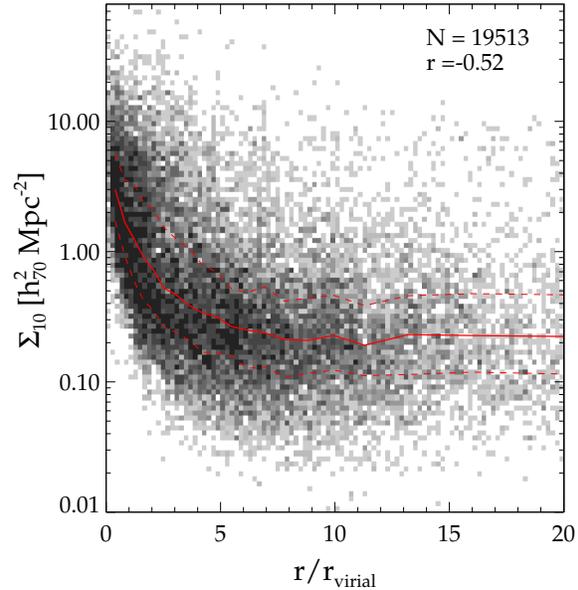}}
\caption{Comparison between local and global definition of environment. The local galaxy
density ($\Sigma_{10}$) is plotted as a function of the cluster-centric
radius ($r/r_{\rm virial}$). Median values of the relation are shown as a solid line, and
their respective quartiles as dashed lines. In the top-right corner is
shown the number of galaxies and the Spearman correlation coefficient for the relation.}
\label{fig:rho_versus_rvirial}
\end{figure}

In this work we use a conventional approach to dealing with galaxy environment.
We adopt a non-parametric method to determine the local number
density of galaxies, based on the $k$th nearest neighbour density
estimator ($k$NN; see, for instance, \citet{fukunaga90} for a statistical
description of such approach). This method fixes a value for $k$ and lets
the volume $V(r)$, centred on a given object and extending to its $k$th
nearest neighbour, be a random variable. This volume is large in low
density regions and small in high density regions. In this way, this 
method provides a spatial description of the density field, which
has been recently explored by \citet{mateus04} in a study of field
galaxies.

When one goes to denser environments, velocity
dispersion is high and neglecting peculiar velocities in distance
estimates could underrate local densities determined with the 
$k$NN method. In order to avoid this effect, a better way to
derive local galaxy density via nearest neighbours is by considering
a projected distribution of galaxies, instead of a spatial one. This
procedure has been intensively used in studies devoted to galaxy
environments, since the classical work by \citet{dressler80} to more
recent works based on 2dFGRS \citep[e.g.][]{lewis02} and
SDSS \citep[e.g.][]{gomez03} data.

A more common approach when dealing with large redshift survey data is to define a limit in redshift space around a given galaxy, for instance in the range $\pm 1000$ km s$^{-1}$, to find its $k$th nearest neighbour \citep[e.g.][]{blanton03a,goto03,balogh04}. This limit emulates a three-dimensional description of the density field by excluding galaxies in the foreground/background with an objective criterion. As noted by \citet{cooper05}, projected density estimates with limits in the line-of-sight velocities between $\pm 1000$ and $\pm 1500$ km s$^{-1}$ are best suited for the description of a broad range of environments.

Here we use the expression $\Sigma_k=k/\pi r_k^{2}$ to estimate the local galaxy density\footnote{Note that in statistics, $k-1$ should replace $k$ in the numerator of the expression for $\Sigma_k$ in order to achieve an unbiased estimate of local density \citep{fukunaga90}.}, where $r_k$ is the projected distance to the $k$th nearest neighbour of a given galaxy. We adopt the usual limit in redshift space of $\pm 1000$~km~s$^{-1}$. Additionally, in our density estimates, we have prevented an incorrect determination of $\Sigma_k$ due to border effects by excluding galaxies whose $k$th neighbours have projected distances greater than the distance of the galaxy to the closest border of the survey region or of the sample volume. Note that in the search of neighbours and in the density estimates we have used all galaxies in the volume limited sample independently of the quality of their spectroscopic data.

We have computed our local galaxy density estimates with the values $k=5$ and $k=10$ in order to assess the influence of the choice of this parameter in the results presented in this work. However, we have found that none of the results discussed in the next sections depend on the values of the $k$ tested here. In the rest of this paper we will adopt the local galaxy density as evaluated from the distance to the tenth neighbour, $\Sigma_{10}$.

We have compared our local density estimates defined by $\Sigma_{10}$ to a global parameter: the distance to the nearest cluster centre from each galaxy in our sample. We searched for galaxy clusters present in our sample with the help of the C4 Cluster Catalogue \citep{miller05}, which contains 748 clusters identified in the spectroscopic sample of the SDSS Data Release 2 (DR2). This catalogue is about 90 per cent complete and 95 per cent pure above $M_{200} = 1 \times 10^{14}$ h$^{-1}$ M$_\odot$ and within $0.03 < z < 0.12$, thus covering the redshift range of our sample. Therefore, for each galaxy in our sample we can compute its distance to the nearest cluster centre, defining in this way a cluster-centric radius normalised by the cluster virial radius. In Fig.~\ref{fig:rho_versus_rvirial}, we show the relation between our density parameter ($\Sigma_{10}$) and the cluster-centric radius ($r/r_{\rm virial}$) only for galaxies located in the regions covered by the DR2 data. There is a clear correlation between the two quantities, mainly for $r/r_{\rm virial} \la 5$, above which we note that the median values of local galaxy density are almost constant.

\begin{figure*}
\centerline{
{\includegraphics[scale=0.7]{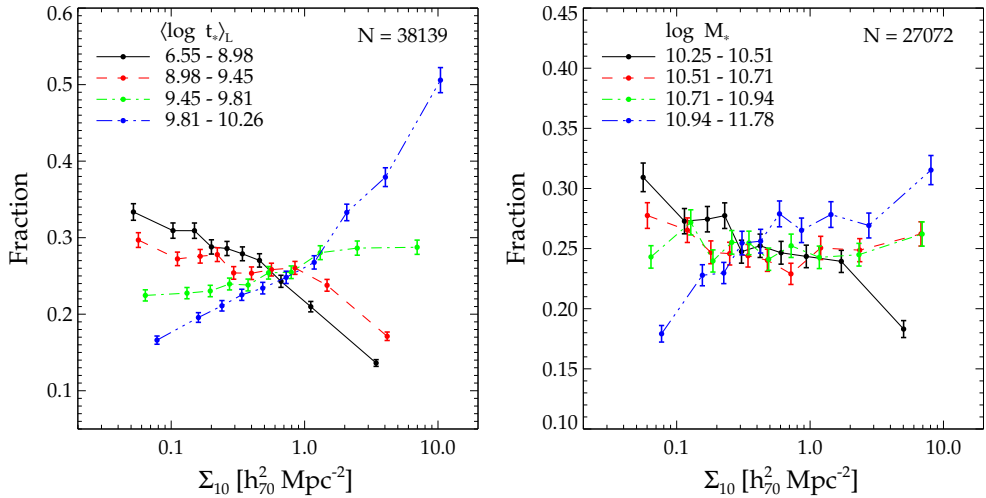}}}
\caption{Left: Fraction of galaxies in bins of light-weighted mean
stellar age (each one containing the same number of galaxies) as a function of the local
galaxy density. Right: The same, but now for galaxies in bins of dust-corrected stellar mass. The numbers at the top-right are the count of galaxies in each panel.}
\label{fig:fraction}
\end{figure*}

A possible source of bias in the estimation of local galaxy density in the SDSS data is due to a technical limitation in the procedure of assignment of fibres to spectroscopic targets closer than 55$\arcsec$ in the sky. This problem, referred as `fibre collision', becomes very important in denser environments and about 8--9 per cent of selected objects are unobserved for this reason \citep{stoughton02}.

In order to assess the significance of this bias in our results we have searched in the DR4 photometric catalogue for galaxies closer than 55$\arcsec$ of each galaxy in our spectroscopic selected sample, which actually have been rejected in the fibre assignment procedure. Note that these missing galaxies only affect our local galaxy density estimate if they have radial velocities close to those of the selected galaxies. However, as they do not have any redshift information we have overestimated the effects of the fibre collision bias by supposing that all such missing `neighbours' are at the same redshift as the selected galaxies. From the total of 63659 selected galaxies in our volume limited sample, we have found about 12 per cent of galaxies with missing `neighbours' closer than 55$\arcsec$. We have then evaluated our density estimates by considering that these missing objects were in fact close neighbours. After redoing our analysis with these new densities we concluded that none of the results shown in this work is affected by the bias due to fibre collision.

\section{The environmental dependence of galaxy properties}

In this section we investigate how some physical properties related to the stellar
populations of nearby galaxies behave as a function of the local galaxy density. We begin by investigating the dependence of the fraction of galaxies of distinct mean stellar ages and masses with the environment, which can be linked to the classical morphology--density relation. The environmental dependence of star-formation activity in galaxies, specially if estimated through the light-weighted mean stellar age parameter, is further investigated. We also analyse how the metallicities of galaxies vary along the density range studied in this work. Finally, the role of stellar mass and luminosity in driving the star-formation--density relation is investigated here, mainly through a principal component analysis applied to our parameter set.

\subsection{Galaxy fractions}\label{sec:fractions}

It is well established that the fraction of galaxies classified
according to distinct morphological types is strongly related with the
environment, with high-density regions typical of galaxy clusters being populated
essentially by elliptical galaxies, in contrast to the high fraction of spirals found
in the field \citep[e.g.][]{dressler80}. This galaxy type segregation is investigated
here in terms of two physical galaxy properties: the mean light-weighted stellar age and
the stellar mass. SEAGal II studied these parameters in light of the
bimodality observed in galaxy populations and found that the mean stellar age of
galaxies is the primary responsible for the dichotomy between
blue and red galaxies seen in the local universe \citep[e.g.][]{strateva01,baldry04}.They also found that the stellar mass has an additional role in the sense that, blue, star forming galaxies are preferentially of low mass.

In Fig.~\ref{fig:fraction} we show how the fraction of galaxies in bins
of light-weighted mean stellar age (left panels) and stellar mass (right panels) correlates with the local galaxy density. Galaxies are grouped in four bins of \tl\ and $\log M_\star/M_\odot$, each one containing the same number of objects. It is clear from Fig.~\ref{fig:fraction} that high-density environments are populated by galaxies with older stellar populations and larger stellar masses, with the fraction of these objects increasing as the environment becomes denser. The transition in galaxy fractions occurs at a local density of $\Sigma^t_{10} \sim 0.7$~$h_{70}^{2}$~Mpc$^{-2}$, above which the population of massive galaxies with older stellar populations begins to dominate. This transition density corresponds, on average, to about 2--3 virial radii from cluster centres, as can be deduced from Fig.~\ref{fig:rho_versus_rvirial}.
Additionally, we note that in the case of $M_\star$, the fractions only change for the two extreme bins, whereas intermediate ones possess constant fractions along $\Sigma_{10}$.
It is worth stressing that the transition shown in Fig.~\ref{fig:fraction} is not the same as the ``break'' at a characteristic galaxy density reported by \citet{lewis02} and \citet{gomez03} that we discuss below, since here we are investigating only galaxy fractions as a function of the local galaxy density, instead of the SFR--density relation itself.

\subsection{Star formation activity}\label{sec:5_2}

Here we revisit the distribution of the \Ha\ equivalent width as a
function of the local galaxy density. In recent years, many works have been
devoted to the study of this relation. For instance, \citet{gomez03} have
investigated the environmental dependence of star formation rate in galaxies
based on the \Ha emission line. Their main result was the finding of a break
at a characteristic local galaxy density resulting in an abrupt decrease of
the EW(\Ha) (or SFR) of galaxies in denser regions. Recently, \citet{tanaka04}
have shown that this break indeed occurs only for fainter galaxies
($M_r^\ast + 1 < M_r < M_r^\ast + 2$), with brighter ($M_r < M_r^\ast + 1$)
ones showing EW(\Ha) only monotonically decreasing with density. Here we use
a more detailed approach to analyse this relation.

\begin{figure*}
\centering\includegraphics[scale=0.7]{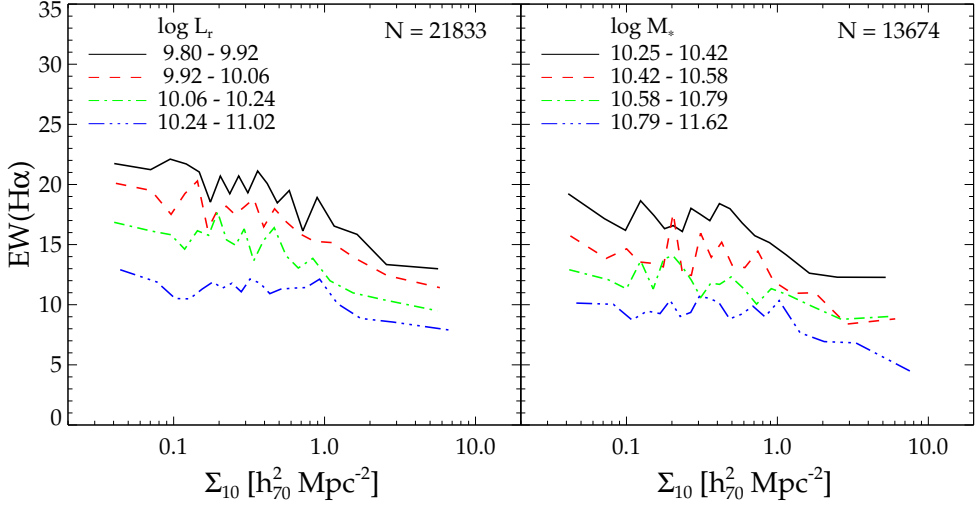}
\caption{Median values of the \Ha equivalent width as a function of local
galaxy density, in bins of galaxy luminosity (left panel) and stellar mass (right
panel). The numbers at the top-right are the count of galaxies in each panel.}
\label{fig:Ha}
\end{figure*}

\begin{figure*}
\includegraphics[scale=0.7]{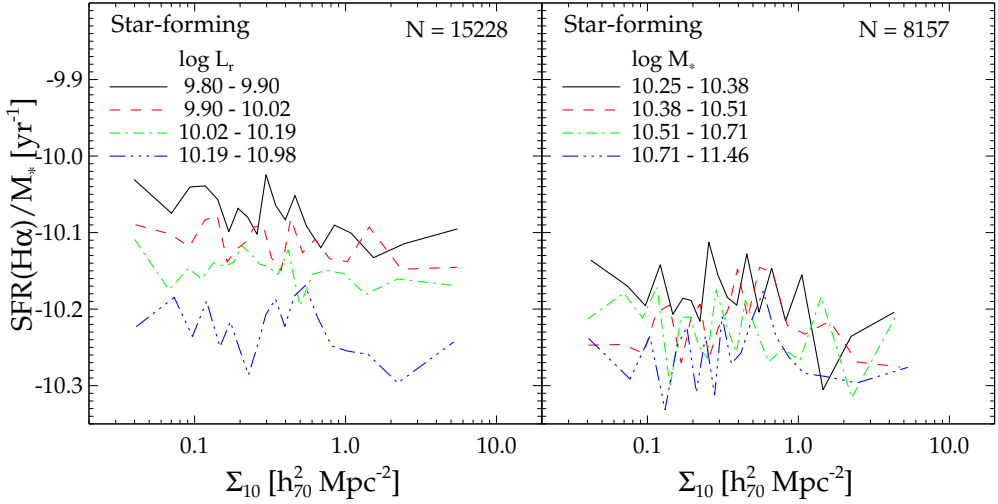}
\caption{Median values of the specific SFR related to the \Ha emission luminosity as a
function of local galaxy density, in bins of galaxy luminosity (left panel) and stellar
mass (right panel), only for galaxies classified as star-forming. The numbers at the top-right are the count of galaxies in each panel.}
\label{fig:SFRHa}
\end{figure*}

In Fig.~\ref{fig:Ha}, we present the median values of EW(\Ha) as a function of local galaxy density. The median values of EW(\Ha) are taken for galaxies grouped in bins of luminosity, $\log L_r/L_\odot$ (left panel) and stellar mass, $\log M_\star/M_\odot$ (right panel). In this figure we show all galaxies with \Ha emission line measured with $S/N$ greater than 3, avoiding to include active nuclei hosts (c.f. Section~\ref{sec:star_formation}). This procedure is very robust since the line measurenments we have done make possible a detailed galaxy classification, as discussed in \citet{mateus06}. Moreover,  with the help of spectral synthesis we can measure only the emission contribution of \Ha line, avoiding to include in our analysis galaxies that would appear with negative values in distributions of EW(\Ha) or SFR(\Ha), as shown in other works (see, for instance, Fig. 4 of \citealt{gomez03}).

We note in Fig.~\ref{fig:Ha} that the median values of EW(\Ha) for
galaxies in all luminosity bins decrease with galaxy density without showing
any break (even if we consider all quartiles of the distribution),
confirming the result found by \citet{tanaka04} for brighter
galaxies (our sample contains galaxies with $M_r < M_r^\star + 1.5$). 
In this sense, the relation between star formation activity and local galaxy
density for bright galaxies is independent of galaxy luminosity, since it
exists in all bins of $\log L_r/L_\odot$. This trend is also seen in the
right panel of Fig.~\ref{fig:Ha}, where galaxies are now divided according to
their stellar masses. The absence of a break in Fig.~\ref{fig:Ha} indicates that
the relation between star formation activity and local galaxy density decreases monotonically over a wide range of environments, as suggested by \citet{mateus04}.

We also investigate this issue by analysing the environmental dependence of the
star formation activity of galaxies spectrally classified as star-forming,
following the same definition as used in \citet{mateus06}. These galaxies have
low stellar masses (about 83 per cent of them have $\log M_\star/M_\odot < 10.67$)
and young mean stellar ages (more than 95 per cent of them have
$\langle\log t_\star\rangle_L < 9.53$). In Fig.~\ref{fig:SFRHa} we show the
specific SFR (SSFR(\Ha)) as a function of the local galaxy density only for the
spectral class of star-forming galaxies. As in Fig.~\ref{fig:Ha}, galaxies are
divided in bins of luminosity (left panel) and stellar mass (right panel).
We note in Fig.~\ref{fig:SFRHa} that the median values of SSFR(\Ha) for star-forming
galaxies have almost constant values along the density bins covered by our sample,
implying that the specific star formation rate of star-forming galaxies does not depend on the local galaxy density, confirming the results found by other investigators
\citep[e.g.][]{balogh04,rines05}.

\begin{figure*}
\centerline{\includegraphics[scale=0.7]{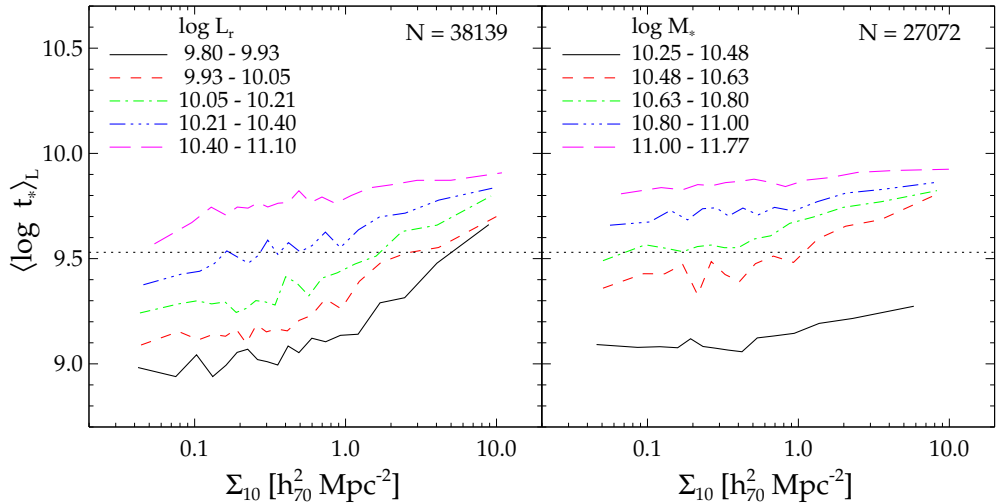}}
\caption{Median values of the mean stellar age weighted by light as a function of the local galaxy density, in bins of galaxy luminosity (left panel) and stellar mass (right panel). The horizontal dashed lines correspond to the optimal value of \tl\ to split galaxies between early and late types. The numbers at the top-right are the count of galaxies in each panel.}
\label{fig:age_rho}
\end{figure*}

\subsubsection{Mean stellar ages}

An important issue concerning to the use of star formation tracers based on emission
line measurements -- like those related to the \Ha emission line -- is the
absence of data for galaxies with little (or even none) recent star formation activity, since the emission lines (if they exist) in the spectra of these
systems are too weak to be measured with appreciable signal-to-noise ratio. Additionally, in galaxies where a starburst is obscured by the presence of huge amounts of dust, the emission lines will also become very weak to be detected.
Thus, results from environmental studies based solely on the EW(\Ha) or
SFR(\Ha) parameters could suffer from this kind of bias.
In Section~\ref{sec:star_formation}, we have shown that the mean
light-weighted stellar age obtained by spectral synthesis has a good
correlation with both EW(\Ha) and SSFR(\Ha),
implying that this parameter is well related to the star formation history of galaxies.
In fact, low values of \tl\ reflect an intense
activity of star formation occurring in a galaxy, whereas high values of it
are associated to most of the galaxy light coming from older stellar
populations. As we measured this parameter for all galaxies in our
sample, it is interesting to investigate its dependence on the environment
defined by the local galaxy density.

In Fig.~\ref{fig:age_rho}, we show the mean stellar age weighted by light 
as a function of the local galaxy density. The median values of \tl\ are taken for galaxies grouped in bins of luminosity and stellar mass.
The behaviour of the mean ages of the stellar populations is distinct when we split
galaxies according to $L_r$ or $M_\star$. In order to clarify these trends in terms of the
bimodal character of the galaxy populations, we also show in this figure, as horizontal dashed lines, the value of \tl\ which better separates early and late galaxy types (see \citet{mateus06} for details).

We note in the left panel of Fig.~\ref{fig:age_rho}, that the median values of \tl\
increase significantly as the environment becomes denser, independently of the galaxy
luminosity, with the relation for fainter galaxies being steeper than that of brighter
ones. The properties of the population of fainter galaxies also tend to change at
$\Sigma_{10} \la 5$~$h_{70}^{2}$~Mpc$^{-2}$, where the median values of \tl\ cross the dividing line which distinguishes galaxies dominated by young stellar populations (late-types) and galaxies with older stars (early-types). On the other hand, the mean stellar ages of galaxies with different stellar masses (seen in the right panel of Fig.~\ref{fig:age_rho}) behave in a distinct way with respect to the local galaxy density. Low-mass galaxies have median values of mean stellar ages slightly varying along the bins of density, showing low values of \tl\ -- characteristic of late-type galaxies -- in all environments. However, this behaviour for the lowest stellar mass bin may be affected by incompleteness due to the low fraction of early-type galaxies selected in the range $10.25 < \log M_\star < 10.48$. The most massive galaxies also show a slight change in the median values of \tl\ as galaxy density increases, but their values are higher everywhere. The steeper relation is seen for galaxies with intermediate stellar masses ($\sim3-6\times10^{10}$~M$_\odot$), which shows a transition from late to early galaxy types at $\Sigma_{10}~\sim~0.3-1.0$~$h_{70}^{2}$~Mpc$^{-2}$.

Therefore, an age--density relation is clearly obtained when galaxies are divided according to their luminosities. However, the relation is not so evident when they are grouped in bins of stellar mass. This indicates that the galaxy luminosity and stellar mass could be playing distinct roles in driving the age--density relation shown in this work. We will appropriately address this question in section \ref{sec:role}.

\begin{figure*}
\centerline{
\includegraphics[scale=0.7]{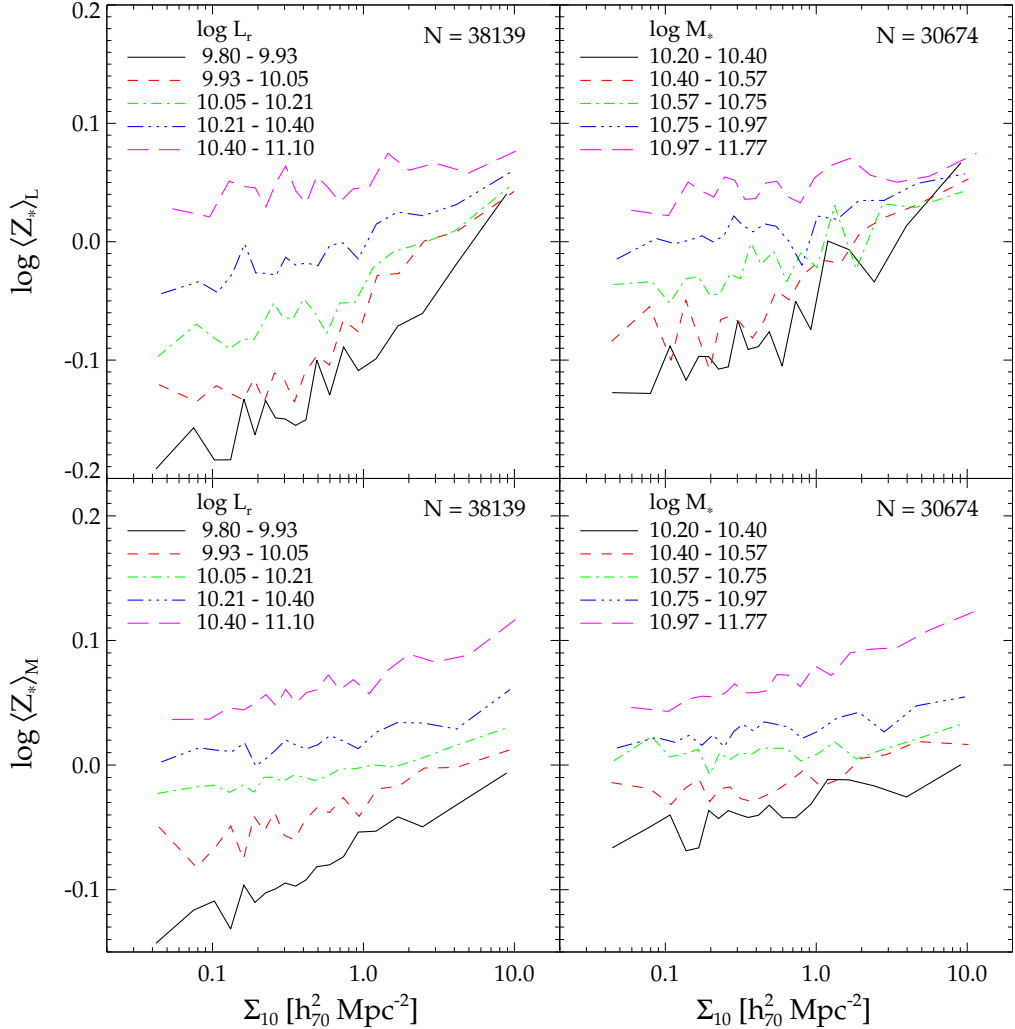}}
\caption{Median values of light-weighted stellar metallicity as a function of local
galaxy density for galaxies in bins of luminosity (left panel) and stellar mass (right
panel). The numbers at the top-right are the count of galaxies in each panel.}
\label{fig:Metallicity}
\end{figure*}

\begin{figure*}
\centerline{
\includegraphics[scale=0.9]{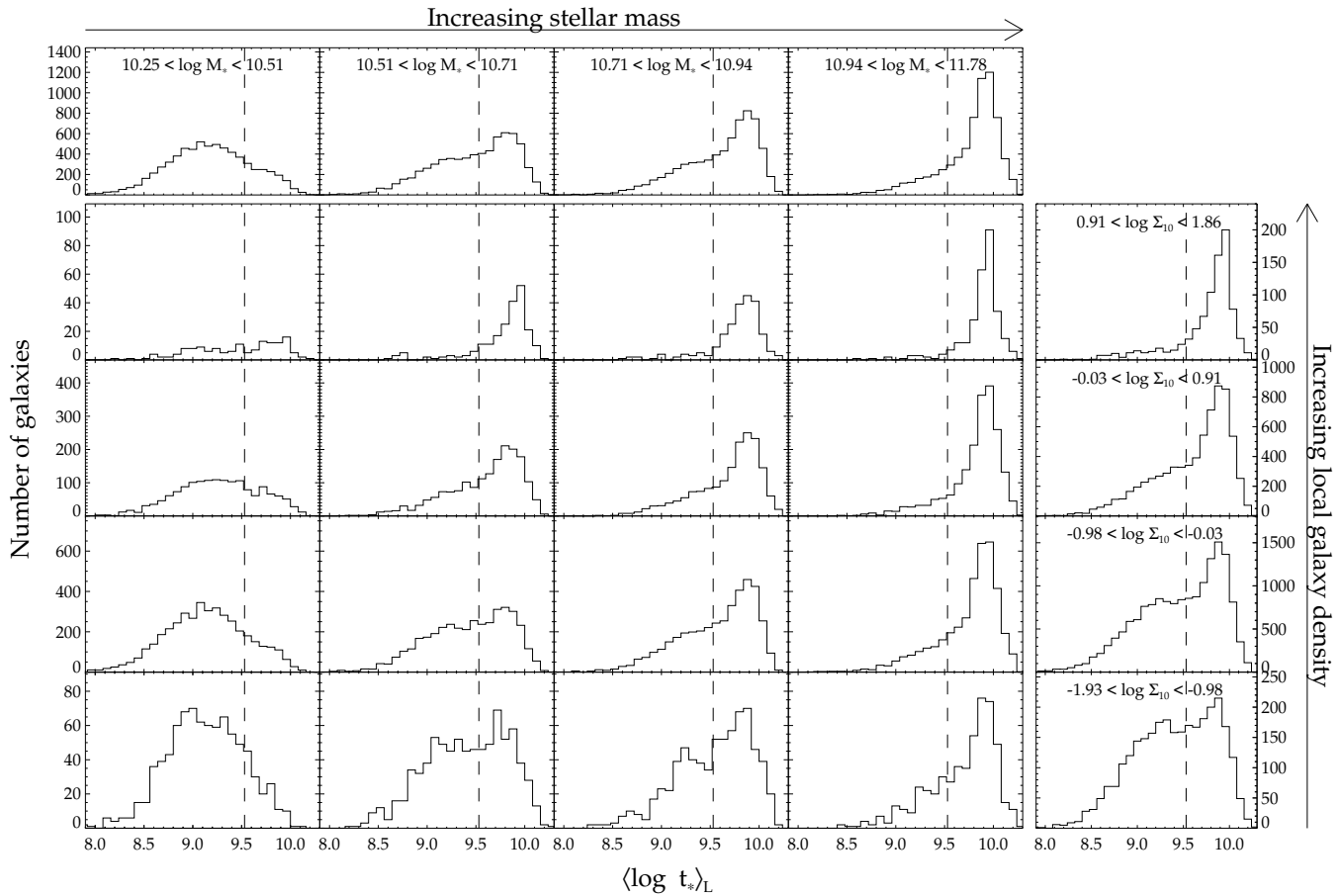}}
\caption{Distributions of mean light-weighted stellar ages for galaxies in bins of
local galaxy density and stellar mass. Dotted lines in each panel are the age value used to distinguish early and late galaxy types. The range for each bin, as well as the cumulative distributions, is shown in top panels, for stellar mass, and in right panels, for local galaxy density.}
\label{fig:histogramas_massa}
\end{figure*}

\begin{figure*}
\centerline{
\includegraphics[scale=0.9]{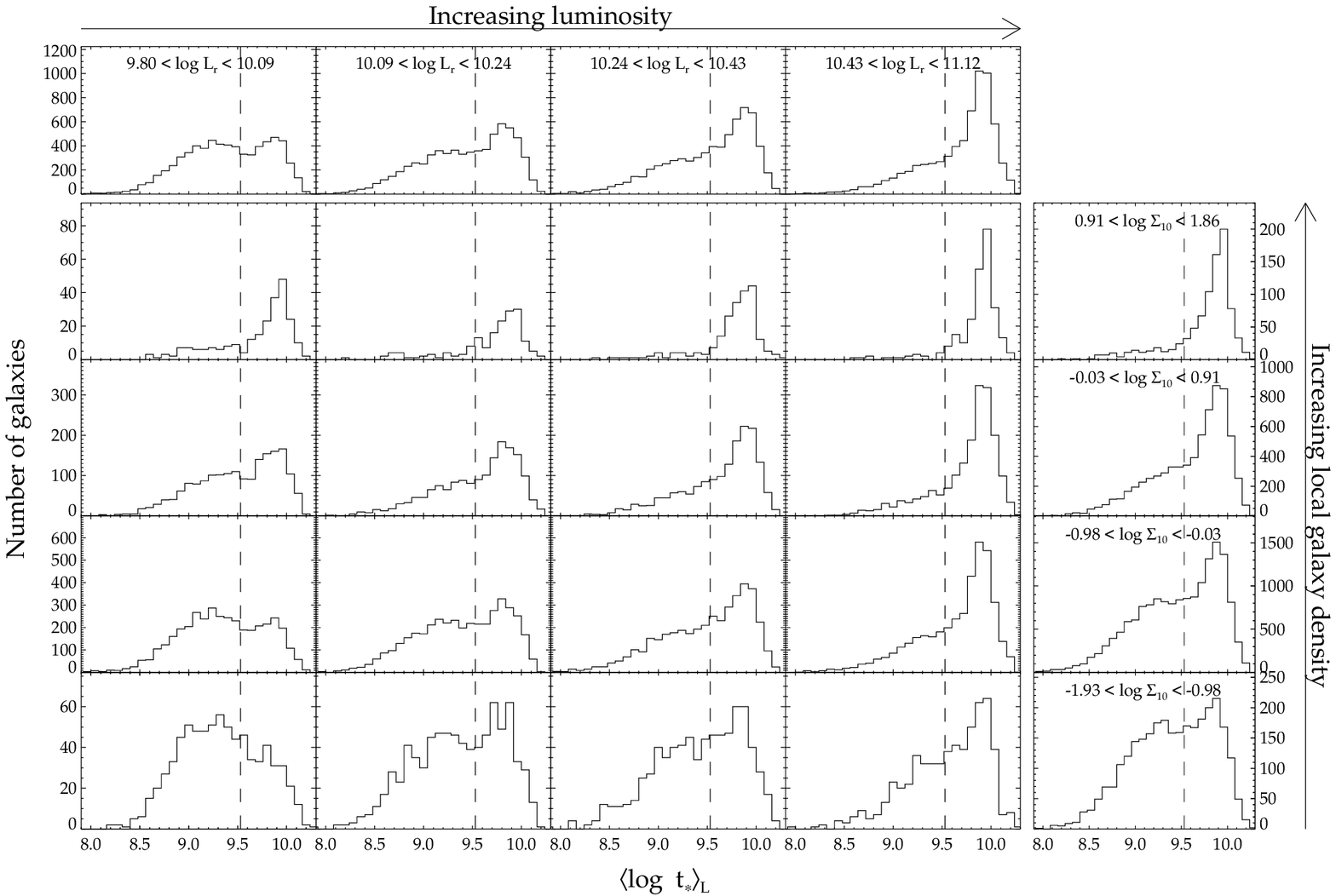}}
\caption{Same as Fig.~\ref{fig:histogramas_massa}, but in bins of luminosity, instead of stellar mass.}
\label{fig:histogramas_luminosidade}
\end{figure*}

\subsection{Stellar metallicities}

We now investigate the environmental dependence of another important physical parameter
of galaxies, the mean stellar metallicity. In Fig.~\ref{fig:Metallicity}, we plot \zl\
and \zm\ versus local galaxy density, for galaxies in bins of luminosity (left panels)
and stellar mass (right panels). The relation between \zl\ and local density becomes less steep as galaxy luminosity increases.
The result of this trend is that fainter galaxies in low-density regions have subsolar
metallicities, whereas in denser regions they have high metallicities similar to that of
the brightest galaxies in our sample. For galaxies divided according to stellar mass, the
trends are similar. The stellar metallicity of low mass galaxies increases as the
environment becomes denser, while massive galaxies have almost constant values of \zl\
along density bins. In the case of the mass weighted mean stellar metallicity (\zm) versus
local galaxy density, the trends are similar, although less steep than those for \zl. The
main difference is that luminous (massive) galaxies also tend to have more metallic stellar populations in denser environments, as shown in the bottom panels of
Fig.~\ref{fig:Metallicity}.

It is interesting to note that the values of \zl\ are related to mean metallicities of
the stellar populations which contribute significantly to galaxy light. Thus, one could
expect that they are directly associated to recent episodes of star formation in
galaxies. In this way, low-mass (or fainter) galaxies in denser environments, which are
currently forming stars (following the results from Fig.~\ref{fig:age_rho}), tend to have
more metal-rich stellar populations than their counterparts in regions of lower
densities. Low mass galaxies have substantial amounts of gas to keep their high star formation rates. This gas could have been pre-enriched earlier preferentially in high-density environments, where the star formation activity was higher in the past \citep[e.g.][]{madau01,scannapieco02}. Thus, one explanation is that these galaxies have already been formed in that denser environments, but have not stopped to form stars until the present epoch. Thus, this result is additionally supported by the observed `downsizing' in galaxy formation \citep[e.g.][among others]{cowie96,kodama04,juneau05}, where massive galaxies have formed most of their stellar masses at early times. However, the processes that have kept the continuous star formation in those low-mass galaxies are not well understood.

\subsection{The role of stellar mass and luminosity}\label{sec:role}

The relation between the mean stellar age of galaxies and their environment (defined
by the local galaxy density) shows that galaxy luminosity and stellar mass play
distinct roles in defining the environmental dependence of star formation properties of
galaxies. In general, that dependence exists independently of $L$ and only for an
intermediate stellar mass range characterised by a transition in galaxy properties.

Another way to visualise these trends is shown in Figs.~\ref{fig:histogramas_massa} and \ref{fig:histogramas_luminosidade}, where we show the distributions of \tl\ for galaxies in vertical bins of local galaxy density, and horizontal bins of stellar mass (Fig.~\ref{fig:histogramas_massa}) and luminosity (Fig.~\ref{fig:histogramas_luminosidade}). Each bin in stellar mass (or luminosity) has been built to have the same number of objects, whereas the bins in $\log \Sigma_{10}$ are equally spaced. From the comparison of these figures, we note that the age distributions are distinct when galaxies are divided according to stellar mass or luminosity, mainly for objects in the extreme bins. In denser environments (corresponding to $0.91 < \log \Sigma_{10} < 1.86$), less luminous galaxies have a \tl\ distribution peaked at older ages, whereas in low-density regions the distribution is double-peaked, unveiling the bimodal character of the galaxy population. On the other hand, in the case of galaxies divided according to stellar mass, the \tl\ distribution for low-mass galaxies in denser environments is broader, without showing any peak. As local density decreases, the distributions are peaked at younger ages, indicative of the dominance of low-mass star-forming galaxies in low-density environments. At the other extreme, represented by very luminous and massive galaxies, the distributions are similar, with the only difference being due to a larger fraction of galaxies with low values of \tl\ in the distributions for luminous galaxies. It is also interesting to note the differences in the distributions for two extreme environments. In denser regions, galaxies divided by luminosity show a conspicuous peak centred at older ages in all bins considered in Fig.~\ref{fig:histogramas_luminosidade}. On the other hand, for galaxies divided by stellar mass there is a growth of the peak at $\langle\log t_\star\rangle_L > 9.5$ from the low-mass to the most massive bin shown in Fig.~\ref{fig:histogramas_massa}. However, note that for the lowest stellar mass bin the absence of older early-type galaxies may be a reflex of our sample selection procedure to overcome the stellar mass incompleteness in our sample (Section~\ref{sec:completeness}).

We also investigate the environmental dependence of a combination of stellar mass and luminosity parameters, namely the mass-to-light ratio of galaxies in our sample as inferred by spectral synthesis. In Fig.~\ref{fig:mass_luminosity}, we show the relation between the galaxy mass-to-light ratio (in $r$-band) and local density in different bins of $r$-band galaxy luminosity for (a) all galaxies in our sample, (b) early-type galaxies with $D_n(4000) > 1.67$  and (c) late-type galaxies with $D_n(4000) < 1.67$. We note that, at fixed luminosity, galaxies inhabiting high-density environments tend to be more massive than their counterparts in regions of low density; the $M_\star/L_r$ ratio of these objects increases by $40-50$ per cent from low to high-density regions. The relation is also steeper for the faintest bin shown in the plot. Additionally, the relation for early and late-type galaxies taken individually are independent on local galaxy density, implying that the increasing of $M_\star/L_r$ as the environment becomes denser is related to the prevalence of massive systems in high-density regions, as shown in Fig.~\ref{fig:fraction}. Only for the most luminous galaxies there is an increment of $M_\star/L_r$ as a function of local density.

\subsection{Principal component analysis}

In the previous sections we have investigated the environmental dependence of the mean light-weighted stellar age, stellar mass, stellar metallicity and mass-to-light ratio in order to understand how galaxy properties are related to the environment. Here we will investigate this issue through a different approach, based on an useful statistical technique frequently adopted to finding patterns in high-dimensional data.

The parameter set discussed in this section was examined in more detail by means of a principal component analysis (PCA) in order to determine the combinations of parameters which summarise the distribution of the whole set. PCA is able to identify patterns in a data set and express the data in such a way as to highlight their similarities and differences. This technique is particularly useful for reducing the dimensionality of a data space by identifying the linear combinations of input parameters with maximum variance, called `principal components' \citep{murtagh87,sodre97}.

A PCA was carried out on the combination of the parameter set $\lbrace\langle\log t_\star\rangle_L, \log M_\star, \log \Sigma_{10}, \log L_r, \log M_\star/L_r\rbrace$ to determine the variables that account for the large amount of variance and hence provide a useful description of the environmental dependence of galaxy properties.
We have followed the same procedure as used in the previous sections by dividing the galaxy sample in bins of luminosity and stellar mass containing the same number of objects. A PCA was then executed for each bin separately.

We found that the principal component of highest variance (PC1) accounts for more than 90 per cent of the total variance for all bins considered. The remaining variance is mainly accounted for the second principal component (PC2). Although each principal component is a linear combination of all the input variables, it is interesting to verify whether some variables in the parameter set are well correlated with the main components because, in this case, they are responsible for a significant fraction of the sample variance. In Table~\ref{table_PCA1} we show the Spearman rank correlation coefficients between the two principal components and the five parameters in our input set, for each bin divided by luminosity and stellar mass. We found that the local galaxy density ($\log \Sigma_{10}$) shows the strongest relation with PC1; the Spearman rank correlation coefficient between this parameter and PC1 is almost unity. Moreover, the PC2 shows a significant correlation with the \tl, with correlation coefficient greater than 0.9. Thus, the plane PC1 $\times$ PC2 is well represented by the age--density relation in all bins of luminosity and stellar mass investigated here.

Additionally, to verify the behaviour of the other parameters we are investigating, we have excluded the mean stellar age from the input set. The PCA was then carried out on the parameter set $\lbrace\log M_\star, \log \Sigma_{10}, \log L_r, M_\star/L_r \rbrace$. For this set, the results indicate that the first principal component alone explains more than 95 per cent of the sample variance for bins divided by luminosity and stellar mass. The second principal component is reponsible for almost all the remaining variance. The Spearman rank correlation coefficients among these two principal components and the parameters in the input set are shown in Table~\ref{table_PCA2} for each bin divided by luminosity and stellar mass. We found that the local galaxy density is again well correlated with the first principal component ($r_S = 1$). The second principal component correlates well with both $M_\star$ and $M_\star/L_r$ for bins in $L_r$ and with $\log L_r$ and $M_\star/L_r$ for bins in $M_\star$.

These results give clear evidence of the SFR--density relation present in our data since the local galaxy density is the primary responsible for the largest amount of sample variance and the mean light-weighted stellar age of galaxies is the secondary source of that variance. Moreover, we note that galaxy luminosity does not have any significant correlation with the principal components, whereas stellar mass and mass-to-light ratio can be considered as the drivers of the sample variance if one excludes the SFR--density relation.

\begin{table}
\centering
\begin{tabular}{lccccc}
\hline \hline
~  & \tl & $\log M_\star$ & $\log \Sigma_{10}$ & $\log L_r$ & $M_\star/L_r$ \\
\hline
\multicolumn{6}{c}{$ 9.80 < \log L_r <  9.96$}\\
PC1  &  0.28 &  0.23 &  0.99 & -0.01 &  0.24\\
PC2  & -0.96 & -0.81 & -0.10 & -0.10 & -0.82\\
\multicolumn{6}{c}{$ 9.96 < \log L_r < 10.13$}\\
PC1  &  0.28 &  0.22 &  1.00 & -0.01 &  0.23\\
PC2  & -0.96 & -0.78 & -0.12 & -0.11 & -0.80\\
\multicolumn{6}{c}{$10.13 < \log L_r < 10.34$}\\
PC1  &  0.23 &  0.18 &  1.00 &  0.03 &  0.19\\
PC2  & -0.97 & -0.72 & -0.10 & -0.15 & -0.75\\
\multicolumn{6}{c}{$10.34 < \log L_r < 11.10$}\\
PC1  &  0.20 &  0.18 &  1.00 &  0.09 &  0.19\\
PC2  & -0.95 & -0.68 & -0.11 & -0.35 & -0.70\\
\hline
\multicolumn{6}{c}{$10.25 < \log M_\star < 10.51$}\\
PC1  &  0.16 &  0.08 &  1.00 & -0.10 &  0.14\\
PC2  & -0.99 & -0.35 & -0.07 &  0.36 & -0.55\\
\multicolumn{6}{c}{$10.51 < \log M_\star < 10.71$}\\
PC1  &  0.24 & -0.01 &  1.00 & -0.17 &  0.17\\
PC2  & -0.97 & -0.08 & -0.09 &  0.47 & -0.53\\
\multicolumn{6}{c}{$10.71 < \log M_\star < 10.94$}\\
PC1  &  0.17 &  0.02 &  1.00 & -0.08 &  0.11\\
PC2  & -0.98 & -0.09 & -0.07 &  0.39 & -0.50\\
\multicolumn{6}{c}{$10.94 < \log M_\star < 11.78$}\\
PC1  &  0.15 &  0.09 &  1.00 &  0.04 &  0.10\\
PC2  & -0.98 & -0.29 & -0.07 & -0.04 & -0.41\\
\hline \hline
\end{tabular}
\caption{Spearman rank correlation coefficients for the five parameters set.}
\label{table_PCA1}
\end{table}

\begin{table}
\centering
\begin{tabular}{lccccc}
\hline \hline
~  & $\log M_\star$ & $\log \Sigma_{10}$ & $\log L_r$ & $M_\star/L_r$ \\
\hline
\multicolumn{5}{c}{$ 9.80 < \log L_r <  9.96$}\\
PC1  &  0.19 &  1.00 & -0.01 &  0.20\\
PC2  & -0.98 & -0.09 & -0.20 & -0.98\\
\multicolumn{5}{c}{$ 9.96 < \log L_r < 10.13$}\\
PC1  &  0.18 &  1.00 & -0.02 &  0.20\\
PC2  & -0.98 & -0.09 & -0.22 & -0.98\\
\multicolumn{5}{c}{$10.13 < \log L_r < 10.34$}\\
PC1  &  0.15 &  1.00 &  0.03 &  0.16\\
PC2  & -0.98 & -0.08 & -0.32 & -0.96\\
\multicolumn{5}{c}{$10.34 < \log L_r < 11.10$}\\
PC1  &  0.17 &  1.00 &  0.09 &  0.18\\
PC2  & -0.99 & -0.10 & -0.79 & -0.69\\
\hline
\multicolumn{5}{c}{$10.15 < \log M_\star < 10.46$}\\
PC1  &  0.07 &  1.00 & -0.10 &  0.14\\
PC2  & -0.24 & -0.08 &  0.87 & -0.96\\
\multicolumn{5}{c}{$10.46 < \log M_\star < 10.67$}\\
PC1  & -0.00 &  1.00 & -0.16 &  0.16\\
PC2  &  0.11 & -0.09 &  0.96 & -0.95\\
\multicolumn{5}{c}{$10.67 < \log M_\star < 10.91$}\\
PC1  &  0.03 &  1.00 & -0.08 &  0.11\\
PC2  &  0.27 & -0.05 &  0.97 & -0.90\\
\multicolumn{5}{c}{$10.91 < \log M_\star < 11.78$}\\
PC1  &  0.10 &  1.00 &  0.04 &  0.10\\
PC2  & -0.92 & -0.03 & -0.95 &  0.07\\
\hline \hline
\end{tabular}
\caption{Spearman rank correlation coefficients for the parameter ser excluding the mean stellar age.}
\label{table_PCA2}
\end{table}

\section{Discussion}

\subsection{Overview}

We have investigated the environmental dependence of some stellar population properties of galaxies in order to advance in our current understanding about the evolution of
galaxies in distinct environments. The most intriguing result shown here is that the relations between local galaxy density and properties related to star
formation is distinct when we split galaxies in bins of luminosity or stellar mass.

The results shown in the last section will be inspected here in the light of two evolutionary paths by which galaxies can evolve. In the first one, galaxy properties (mainly related to star formation and gas properties) are affected by environment through well known physical mechanisms acting on galaxies. This path, linked directly to the environment, gives origin to a \textit{nurture} perspective for galaxy evolution. The second path is related to the initial conditions established during galaxy formation, which could account for the relations between galaxy properties and environment. Thus, it is related to a \textit{nature} perspective driving galaxy evolution.

\begin{figure*}
\resizebox{\textwidth}{!}{\includegraphics{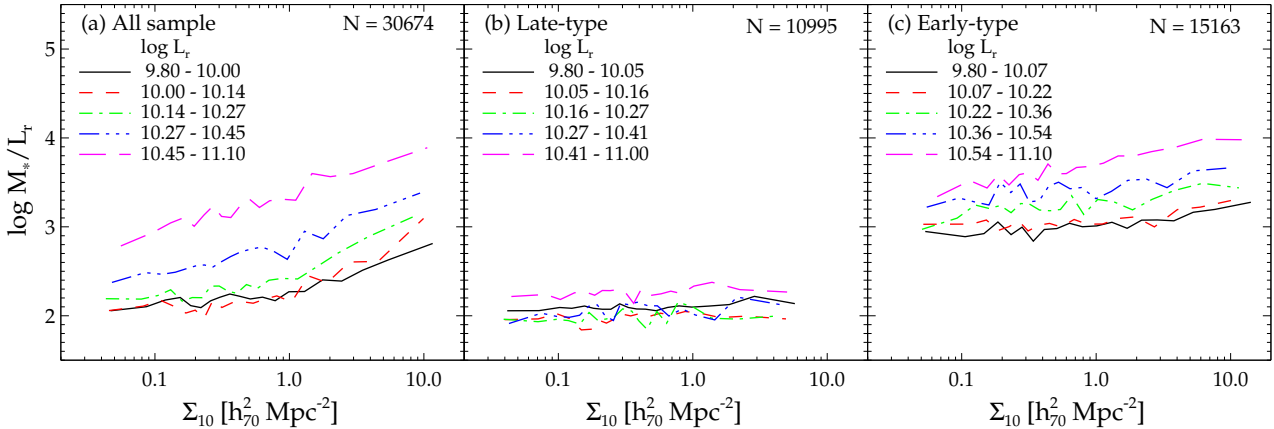}}
\caption{Median values of the mass-to-light ratio as a function of local galaxy density
for galaxies in bins of luminosity. The plot in (a) is for all galaxies in our sample,
(b) for early-type galaxies with $D_n(4000) > 1.67$ and (c) for late-type galaxies with
$D_n(4000) < 1.67$. The numbers at the top-right are the count of galaxies in each panel.}
\label{fig:mass_luminosity}
\end{figure*}

\subsection{The star-formation--density relation: a nurture perspective?}

Recent galaxy redshift surveys have improved our understanding about what is
happening with the star formation in galaxies inhabiting different
environments. Works carried out by \citet{lewis02} using the 2dFGRS data,
and \citet{gomez03} using the EDR SDSS data, have revealed a
characteristic density at $\sim 1$~$h_{75}^{2}$~Mpc$^{-2}$ (corresponding to a
cluster-centric radius of 3--4 virial radii) associated to a ``break'' in the
SFR-density relation. Below this density the SFR increases only slightly,
whereas at denser regions it is strongly suppressed. \citet{tanaka04}
have complemented these studies by finding that the break in the SFR-density
relation is seen only for fainter galaxies
($M_r^\ast + 1 < M_r < M_r^\ast + 2$), while the relation for brighter
galaxies ($M_r < M_r^\ast + 1$) shows no break. Various physical mechanisms are advocated  to explain these trends, but in general it is supposed that star formation properties of bright galaxies are affected by low-velocity interactions, while those of faint galaxies are affected by starvation (or strangulation).

In this work we have investigated relatively bright objects ($M_r \le M_r^\ast + 1.5$) and for a set of galaxies with \Ha line in emission we have found no break or characteristic density in the relation between EW(\Ha) and local galaxy density, independently of galaxy luminosity (see Fig. \ref{fig:Ha}). Apparently, this result is in conflict with the conclusions reached by \citet{lewis02}, \citet{gomez03}, and \citet{tanaka04} that the star formation properties of (faint) galaxies change abruptly when one goes to denser environments. Here we clarify this question by investigating the colour--density relation for galaxies in our sample. For comparison purposes, this procedure is similar to that done by \citet{tanaka04} and shown in their Fig.~2. It is worth stressing that galaxy colours are measured in an uniform way by the SDSS photometric pipeline, that is, they do not depend on the methodology employed to compute them. In addition, the colour--density relation, as discussed by \citet{tanaka04}, is very similar to the EW(\Ha)--density relation.

\begin{figure*}
\includegraphics[scale=0.7]{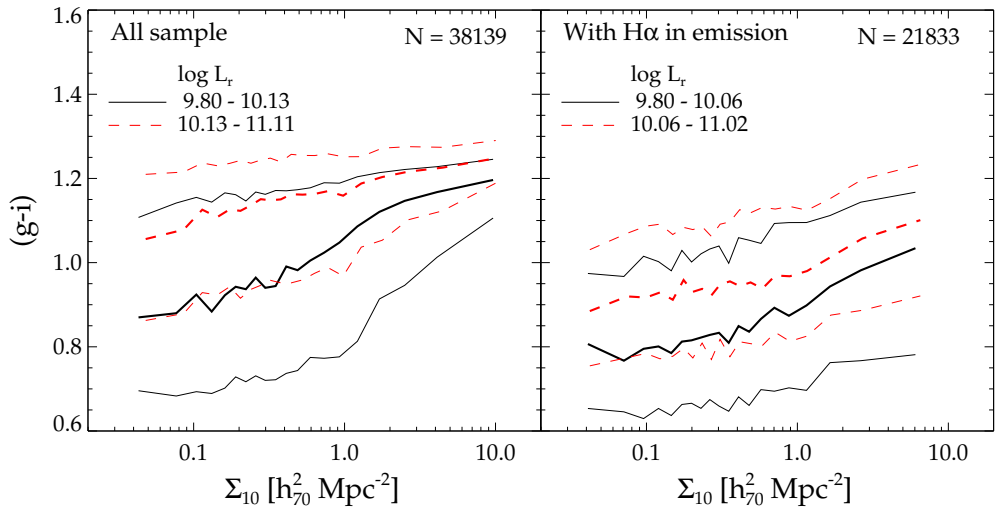}
\caption{Left panel: Median values (thick lines) and respective quartiles (thin lines) of $(g-i)$ colour as a function of local galaxy density for all galaxies in our sample divided in two bins of galaxy luminosity. Right panel: The same, but now restricted only to galaxies with \Ha detected in emission. The numbers at the top-right are the count of galaxies in each panel. }
\label{fig:colour-density}
\end{figure*}

We show in Fig.~\ref{fig:colour-density} the relation between the $(g-i)$ colour and the local galaxy density for galaxies divided in two bins of $r$-band luminosity. In the left panel, we show the relation for all galaxies in our sample. A clear break is seen at $\Sigma_{10} \sim 1$~$h_{70}^{2}$~Mpc$^{-2}$ in the relation for faint galaxies, in complete agreement with the results discussed by \citet{tanaka04}. On the other hand, bright galaxies tend to become redder in dense environments following a monotonically increasing relation. In the right panel of Fig.~\ref{fig:colour-density} we show the same relation but now restricting the sample to a set of galaxies with \Ha detected in emission (avoiding to include AGN hosts, as discussed in Section~\ref{sec:star_formation}). In this case, galaxies also tend to become redder in denser environments, but without showing any break at a characteristic density, even for faint galaxies. Thus, from the inspection of Fig.~\ref{fig:colour-density} we conclude that the break seen in the colour--density relation, or in the SFR--density relation discussed by \citet{gomez03}, is associated to the reduced fraction of star-forming galaxies in denser environments, as discussed in Section~\ref{sec:fractions}, confirming the results shown by \citet{balogh04}. Thus, in \citet{gomez03} and \citet{tanaka04}, for instance, the inclusion of absorption-line dominated galaxies (those with negative EW values) originated the break in the star formation--density relation at a particular local density, which actually is related to an environment where the fraction of galaxies without recent star formation activity begins to dominate.

These results also indicate that the decrease in the fraction of star-forming galaxies in dense environments occurs in a wide range of densities. A similar conclusion was reached by \citet{mateus04} in a study on the environmental dependence of the fraction of star-forming and passive galaxies in a sample of field galaxies drawn from the 2dFGRS data. Even in rarefied environments the fraction of star-forming galaxies decreases significantly with increasing local galaxy density. Thus, there is a general consensus that physical mechanisms suppressing star formation in galaxies are not inherent only to the densest environments related to galaxy clusters. It is also interesting to note that faint and bright galaxies seem to have evolved through distinct evolutionary paths, and different physical mechanisms could be at work.

However, the relevance of such mechanisms acting on galaxies today may be important
only in some particular cases. \citet{balogh04} have confirmed this trend
by combining samples from the 2dFGRS and SDSS, with emphasis on environments
related to galaxy groups. These authors argue that galaxies in dense regions
have had typically more interactions than those in low-density regions over a
longer period of time. Thus, it seems that galaxy transformations induced
by environment have taken place more effectively at higher redshifts. 
In fact, \citet{kauffmann04} have demonstrated that relations between
structural parameters and stellar mass have little dependence on local galaxy
density, indicating that these relations were already established at early
times. In this way, mechanisms that are effective in disturbing the structure of
galaxies (affecting the concentration index and surface mass density) are not
favoured by these results.

\subsection{A natural path for galaxy evolution}

In the last section, we have shown that the nurture hypothesis which is
invoked to explain galaxy evolution in distinct environments also explains the
observed star formation-density relation in the local universe. Indeed, in the
hierarchical scenario of galaxy formation, as galaxy clustering evolves the density
around a galaxy tends to increase in all environments. Higher density probably means more
interactions which could, in principle, play a significant role in originating the
relation between star formation activity in galaxies and local galaxy density.

The first numerical simulations with  CDM initial
conditions \citep[e.g.][]{frenk85, frenk88} already suggested that the
morphology--density relation is a natural product of hierarchical clustering. 
The high-redshift progenitors of today cluster galaxies were formed in
high-density regions, which tend to collapse earlier than low-density
environments \citep{kaiser84, davis85, bardeen86}. Galaxies formed in
high-density regions tend to be ellipticals, because disks are destroyed
by frequent interactions and the hot gas does not cool enough to form
new stars. Thus, it is expected that the oldest galaxies were
formed earlier and preferentially in high-density regions; galaxies in
low-density environments are able to keep their cold gas and form stars. 
Consequently, a natural path for galaxy evolution emerges, 
with galaxies in denser environments being more evolved than those in 
low-density regions \citep{benson01}.

Recently, \citet{juneau05}, using  a sample of high-redshift galaxies 
from the Gemini Deep Deep Survey (GDDS), investigated the star 
formation history of galaxies and its dependence on stellar mass. 
They showed that the most massive galaxies formed most of their stars 
earlier than intermediate and low-mass galaxies. This trend gives 
support to an apparent anti-hierarchical behaviour, commonly 
referred as `downsizing' in galaxy formation \citep[e.g.][]{cowie96,kodama04}. 
This picture, where massive galaxies stopped star formation at early times, 
whereas most low mass systems continue to form stars actively, 
is supported by the analysis of galaxies in the local universe 
\citep{kauffmann03a, heavens04, mateus06}, as well as by the 
observation of a large population of massive galaxies at 
high-redshift \citep[e.g.][among others]{glazebrook04,chen04}.

These trends have been recovered by recent high-resolution simulations
\citep[e.g.][]{weinberg04} and semi-analytic models of galaxy formation
\citep[e.g.][]{menci05,delucia05}. On the observational side, the discovery of an
overdensity of galaxies (or proto-cluster) at $z = 2.3$ by \citet{steidel05}, with a
significant fraction of old galaxies, is consistent with the theoretical expectation for
the acceleration of structure formation in denser regions. This is in agreement with
the conclusions drawn by \citet{einasto05}, that clusters in high-density environments
evolve more rapidly than those in low-density regions.

\subsection{A nature via nurture scenario}

In this work, we have confirmed that high-density environments are dominated by massive galaxies with the oldest stellar populations in the local universe, which can be associated to the early-type galaxies of the morphology-density relation. We have also found that the recent star formation activity of less massive galaxies, as inferred by the mean light-weighted age of their stellar populations, does not change significantly along the density range presented in this work. This result suggests that these low-mass galaxies have large amounts of star formation in low as well as in high-density regions of the local universe. However, as the lowest stellar mass bin considered in this work is dominated by late-type galaxies, with the fraction of low-mass early-type galaxies being affected by the selection criteria adopted here, this result should be confirmed by other observations. At the other extreme, the most massive galaxies in our sample have high values of mean stellar age independently of the environment where they are found. In short, massive galaxies have older stellar populations everywhere. Only for an intermediate mass range ($\sim 3-6 \times 10^{10}$ M$_\odot$), associated to the transition in galaxy properties \citep[see, for instance, ][]{mateus06}, the recent star formation activity of galaxies appears to decrease with increasing galaxy density. On the other hand, when we take galaxies divided according to their luminosities, instead of stellar masses, the star formation activity of brighter galaxies and of fainter ones decreases significantly when one goes from low to high-density environments, resulting in the well know star formation--density relation for galaxies selected by luminosity \citep[e.g.][]{lewis02, gomez03, balogh04, tanaka04, rines05}.

We have also shown in Fig.~\ref{fig:mass_luminosity} that, on average, galaxies with similar luminosities in high-density environments have higher mass-to-light ratios, that is, they tend to be more massive than their counterparts in regions of low density (see also \citealt{tully05} for a similar result related to galaxy groups, and \citealt{einasto05} for galaxy clusters). Thus, the fact that galaxies of all luminosities present star formation activity decreasing with increasing density is related to the higher stellar masses of galaxies in denser environments, independently of their luminosities. Additionally, from recent results concerning the existence of a downsizing in the processes regulating star formation in galaxies, we expect that most massive galaxies have stopped to form stars at early times, perhaps thanks to internally driven mechanisms such as a mass threshold below which star formation was not successful \citep{martin01,jimenez05}. Thus, the star formation--density relation found for galaxies divided according to their luminosities, being steeper for fainter galaxies \citep{tanaka04}, should be a ``natural'' consequence of galaxy evolution.

These results are also supported by the recent findings obtained by \citet{tanaka05} in a study on the build up of the colour-magnitude relation of galaxies as a function of the environment. They also confirm the downsizing way for the evolution of galaxy properties, with the star formation processes being displaced from massive to low-mass galaxies, and from galaxies in denser to low-density environments, as the evolution proceeds. In addition, our work can also be related to the recent study carried out by \citet{poggianti05}, who have investigated the evolution of the proportion of star-forming galaxies (defined with the EW \oii) in clusters since $z = 0.4-0.8$ (from a sample obtained with the ESO Distant Cluster Survey) to $0.04 < z < 0.08$ (from a SDSS cluster sample). Poggianti et al. argue that the star formation-density relation should have a `primordial' component and an `evolved' one, each of them showing distinct environmental dependences, and that the relation is established at very high redshift at the moment of formation of the first stars in galaxies. This scenario is particularly linked to the trends shown by our own results.

The trend related to stellar metallicity shown in Fig.~\ref{fig:Metallicity} reinforces these ideas: low-mass galaxies in denser environments are metal-rich compared with those in low-density regions. This arises from the fact that denser regions tend to be more evolved than rarefied environments. Thus, low-mass galaxies in dense regions have already been established in such environments, but have not stopped to form stars until the present epoch, contrarily to the evolution of massive systems. In this sense, a natural path for galaxy evolution proceeds via a nurture way mainly at high-redshifts: massive galaxies have been formed in denser regions and evolved in an accelerated way, contrasting with a more \textit{unsocial} life of low-mass galaxies preferentially inhabiting low density regions of the universe. This is also related to the timing of the process of formation of early-type galaxies that occurred about 1--2 Gyr earlier in cluster environments compared to the field \citep{thomas05,clemens06}. In this view, denser environments behave as an accelerator of the galaxy formation process, amplifying small natural differences imprinted on galaxies in their formation epoch. This results that ``nature'' necessarily acts \textit{via} ``nurture'' effects.

In this \emph{nature via nurture} scenario for galaxy evolution, the environmental dependence of star formation properties of galaxies, commonly related to some (or various) physical mechanisms acting on late-type galaxies infalling onto galaxy clusters, is determined by the higher $M_\star/L$ ratio of galaxies in denser environments, in addition to the observed downsizing in galaxy formation. These two ingredients play a crucial role in defining the star formation-density relation observed in luminosity divided samples.
Nevertheless, we note that environmental effects on galaxy properties really take place in some cases, such as the existence of post-starburst galaxies in clusters \citep[e.g.][]{dressler83,dressler99,poggianti99,tran03}, passive spirals \citep[e.g.][]{goto03}, and short starburst galaxies \citep{balogh99, mateus04}. However, the fraction of such `environment-product' objects is very reduced, indicative of a very rapid and efficient mechanism acting on these galaxies or actually due to the rarity of such transformations in the nearby universe.

\section{Summary}

In this fourth paper of the SEAGal collaboration, we have shed some light on the discussion about the environmental dependence of galaxy properties in the local universe.
We based our analysis on the stellar population properties of galaxies in a volume-limited sample drawn from the SDSS Data Release 4. The application of a spectral synthesis method to the data produces robust estimators for mean stellar ages, mean stellar metallicities and stellar mass, which have been used in this work to characterise the stellar populations of galaxies. The environment is defined by the local galaxy density estimated from a nearest neighbour approach, and by the distance to cluster centres obtained from a public catalogue of SDSS clusters. The main approach used in this study is the comparison of the relations between galaxy properties and environment for galaxies divided in intervals of luminosity and stellar mass. We summarise our main findings below:

\begin{enumerate}
\item We recover the star formation--density relation in terms of the mean light-weighted stellar age, \tl, which is strongly correlated with star formation parameters derived from \Ha emission line.

\item We confirm that high-density environments are populated by a large fraction of massive galaxies with old stellar populations, in opposition to low-density regions, dominated by low-mass galaxies actively forming stars. We also note that galaxies with intermediate stellar masses have constant fractions along the range of densities covered by our sample. The transition in galaxy fraction occurs at $\Sigma^t_{10} \sim 0.7$~$h_{70}^{2}$~Mpc$^{-2}$, corresponding to 2--3 virial radii from cluster centres.

\item The environmental dependence of \tl\ is distinct when we divide galaxies according to luminosity or stellar mass. The relation between mean age and local density is remarkable for galaxies in all bins of luminosity. On the other hand, only for a intermediate stellar mass interval (associated to a transition in galaxy properties) the relation shows a change in galaxy properties. For low-mass galaxies, the relation slightly increases along the density range, but showing low values of \tl\ -- characteristic of late-type galaxies -- in all environments. The most massive galaxies also show a slight change in the median values of \tl\ as galaxy density increases, but their values are higher everywhere.

\item We find that the distinct behaviours of the \emph{age--density relation} for galaxies divided by luminosity or stellar mass are associated to the large stellar mass of galaxies with the same luminosity in dense environments. In other words, the well known star formation--density relation results from the prevalence of massive systems in high-density environments, independently of galaxy luminosity, in addition to the observed downsizing in galaxy formation, where the star formation is shifted from massive galaxies at early times to low-mass ones as the universe evolves.

\item A principal component analysis of our parameter set reveals that the local galaxy density is the primary responsible for the total variance present in our data, whereas the mean light-weighted stellar age of galaxies is the second one. This result reflects that the age--density relation is the main driver of the environmental dependence observed in some galaxy properties.

\item The mean stellar metallicity of less massive/luminous galaxies increases in high-density regions, indicating that even low-mass galaxies in dense environments tend to be more evolved than their counterparts in low-density regions.

\item Our results support that a natural path for galaxy evolution proceeds \textit{via} a nurture way mainly at high-redshifts: massive galaxies have been formed in denser regions and evolved in an accelerated way, contrasting with a more \textit{unsocial} life of low-mass galaxies preferentially inhabiting low density regions of the universe.

\end{enumerate}

In this work, the concept of `nature via nurture' \citep[see ][for a genetic view of this expression]{ridley03} in the scenario of galaxy evolution was advocated in order to summarise the results obtained here related to the environmental dependence of galaxy properties in the local universe. Many steps are needed towards a comprehensive view of the processes which have guided galaxy formation and evolution, mainly those related to the properties of the stellar content of galaxies in high redshifts. We expect to contribute further to these issues in other papers of this series on Semi-Empirical Analysis of Galaxies.

\section*{Acknowledgements}
We thank the anonymous referee for comments and suggestions
that helped improve the paper. We thanks financial support from CNPq,
FAPESP and the France-Brazil PICS 
program. All the authors wish to thank the team of the
Sloan Digital Sky Survey (SDSS) for their dedication to a project 
which has made the present work possible.

Funding for the Sloan Digital Sky Survey has been provided by the 
Alfred P. Sloan Foundation, the Participating Institutions, the 
National Aeronautics and Space Administration, the National Science 
Foundation, the U.S. Department of Energy, the Japanese 
Monbukagakusho, and the Max Planck Society. 
The SDSS is managed by the Astrophysical Research Consortium (ARC) 
for the Participating Institutions. The Participating Institutions are 
The University of Chicago, Fermilab, the Institute for Advanced Study, 
the Japan Participation Group, The Johns Hopkins University, the 
Korean Scientist Group, Los Alamos National Laboratory, the 
Max-Planck-Institute for Astronomy (MPIA), the Max-Planck-Institute 
for Astrophysics (MPA), New Mexico State University, University of 
Pittsburgh, University of Portsmouth, Princeton University, the 
United States Naval Observatory, and the University of Washington.

\bsp

\label{lastpage}


\begin{thebibliography}{}

\bibitem[\protect\citeauthoryear{Adelman-McCarthy et al.}{2006}]{DR4} Adelman-McCarthy J.~K., et al., 2006, ApJS, 162, 38

\bibitem[\protect\citeauthoryear{Baldry et~al.}{2004}]{baldry04}
Baldry I.~K., Glazebrook K., Brinkmann J., Ivezi\'{c} Z., Lupton R.~H., Nichol R.~C., Szalay A.~S., 2004, ApJ, 600, 681

\bibitem[\protect\citeauthoryear{Balogh et al.}{1999}]{balogh99}
Balogh M.~L., Morris S.~L., Yee H.~K.~C., Carlberg R.~G., Ellingson E., 1999, ApJ, 527, 54

\bibitem[\protect\citeauthoryear{Balogh et al.}{2004}]{balogh04}
Balogh M.~L., Baldry I.~K., Nichol R., Miller C., Bower R., Glazebrook K., 2004, ApJL, 615,101

\bibitem[\protect\citeauthoryear{Bardeen et al.}{1986}]{bardeen86}
Bardeen J.~M., Bond J.~R., Kaiser N., Szalay A.~S., 1986, ApJ, 304, 15

\bibitem[\protect\citeauthoryear{Bekki, Couch \& Shioya}{2001}]{bekki01a}
Bekki K., Couch W.J., Shioya Y., 2001, PASJ, 53, 395

\bibitem[\protect\citeauthoryear{Benson et al.}{2001}]{benson01}
Benson A.~J., Frenk C.~S., Baugh C.~M., Cole S., Lacey C.~G., 2001, MNRAS, 327, 1041  

\bibitem[\protect\citeauthoryear{Blanton et al.}{2003a}]{blanton03a}
Blanton M.~R., et al., 2003a, ApJ, 594, 186

\bibitem[\protect\citeauthoryear{Blanton et al.}{2003a}]{blanton03b}
Blanton M.~R., Lin H., Lupton R.~H., Maley F.~M., Young N., Zehavi I., Loveday J., 2003b, AJ, 125, 2276

\bibitem[\protect\citeauthoryear{Blanton et al.}{2003b}]{blanton03kcorrect}
Blanton M.~R, Brinkmann J., Csabai I., Doi M., Eisenstein D., Fukugita M., Gunn J.~E., Hogg D.~W., Schlegel D.~J., 2003c, AJ, 125, 2348 

\bibitem[\protect\citeauthoryear{Bravo-Alfaro et al.}{2000}]{bravo00}
Bravo-Alfaro H., Cayatte V., van Gorkon J.H., Balkowski C., 2000, AJ, 119, 580

\bibitem[\protect\citeauthoryear{Brinchmann \& Ellis}{2001}]{brinchmann01}
Brinchmann J., Ellis R.~S., 2000, ApJ, 536, L77  

\bibitem[\protect\citeauthoryear{Bruzual \& Charlot}{2003}]{bc03}
Bruzual G., Charlot S., 2003, MNRAS, 344, 1000

\bibitem[\protect\citeauthoryear{Bundy et al.}{2006}]{bundy06}
Bundy K., et al., 2006, submitted (astro-ph/0512465)

\bibitem[\protect\citeauthoryear{Cardelli, Clayton, \& Mathis}{Cardelli et al.}{1989}]{cardelli89}
Cardelli J.~A., Clayton G.~C., Mathis J.~S., 1989, ApJ, 345, 245

\bibitem[\protect\citeauthoryear{Carter et al.}{2001}]{carter01}
Carter B.~J., Fabricant D.~G., Geller M.~J., Kurtz M.~J., McLean B., 2001, ApJ, 559, 606

\bibitem[\protect\citeauthoryear{Cen \& Ostriker}{1993}]{cen93}
Cen~R., Ostriker~J.~P., 1993, ApJ, 417, 415

\bibitem[\protect\citeauthoryear{Chen \& Marzke}{2004}]{chen04}
Chen H.-W., Marzke R.~O., 2004, ApJ, 615, 603

\bibitem[\protect\citeauthoryear{SEAGal I}{}]{seagal1}
Cid Fernandes R., Mateus A., Sodr\'e L., Stasi\'nska G., Gomes J.~M, 2005, MNRAS, 358, 363

\bibitem[\protect\citeauthoryear{Clemens et al.}{2006}]{clemens06} Clemens M.~S., Bressan A., Nikolic B., Alexander P., Annibali F., Rampazzo R., 2006, MNRAS, 370, 702

\bibitem[\protect\citeauthoryear{Cooper et al.}{2005}]{cooper05}
Cooper M.~C., Newman J.~A., Madgwick D.~S., Gerke B.~F., Yan R., Davis M., 2005, ApJ, 634, 833  

\bibitem[\protect\citeauthoryear{Cowie et al.}{1996}]{cowie96}
Cowie L.~L., Songaila A., Hu E.~M., Cohen J.~G., 1996, AJ, 112, 839

\bibitem[\protect\citeauthoryear{Davis et al.}{1985}]{davis85}
Davis M., Efstathiou G., Frenk C.~S., White S.~D.~M., 1985, ApJ, 292, 371  

\bibitem[\protect\citeauthoryear{de Lucia et al.}{2006}]{delucia05}
De Lucia G., Springel V., White S.~D.~M., Croton D., Kauffmann G., 2006, MNRAS, 366, 499  

\bibitem[\protect\citeauthoryear{Dressler}{1980}]{dressler80}
Dressler A., 1980, ApJ, 236, 351 

\bibitem[\protect\citeauthoryear{Dressler \& Gunn}{1983}]{dressler83}
Dressler A., Gunn J.E., 1983, ApJ, 270, 7

\bibitem[\protect\citeauthoryear{Dressler et al.}{1999}]{dressler99}
Dressler A., Smail I., Poggianti B.~M., Butcher H., Couch W.~J., Ellis R.~S., Oemler A.~J., 1999, ApJS, 122, 51  

\bibitem[\protect\citeauthoryear{Einasto et al.}{2005}]{einasto05}
Einasto J., Tago E., Einasto M., Saar E., Suhhonenko I., Hein{\"a}m{\"a}ki P., H{\"u}tsi G., Tucker D.~L., 2005, A\&A, 439, 45

\bibitem[\protect\citeauthoryear{Frenk et al.}{1985}]{frenk85}
Frenk C.~S., White S.~D.~M., Efstathiou G., Davis M., 1985, Nature, 317, 595

\bibitem[\protect\citeauthoryear{Frenk et al.}{1988}]{frenk88}
Frenk C.~S., White S.~D.~M., Davis M., Efstathiou G., 1988, ApJ, 327, 507

\bibitem[\protect\citeauthoryear{Fujita \& Nagashima}{1999}]{fujita99}
Fujita Y., Nagashima M., 1999, ApJ, 516, 619

\bibitem[\protect\citeauthoryear{Fukunaga}{1990}]{fukunaga90}
Fukunaga K., 1990, Introduction to Statistical Pattern Recognition, 2nd edn. Academic Press

\bibitem[\protect\citeauthoryear{Gavazzi et al.}{2002}]{gavazzi02}
Gavazzi G., Boselli A., Pedotti P., Gallazzi A., Carrasco L., 2002, A\&A, 396, 449

\bibitem[\protect\citeauthoryear{Glazebrook et al.}{2004}]{glazebrook04}
Glazebrook K., et al., 2004, Nature, 430, 181

\bibitem[\protect\citeauthoryear{G\'omez et al.}{2003}]{gomez03}
G\'omez P.~L., et al., 2003, ApJ, 584, 210

\bibitem[\protect\citeauthoryear{Goto et al.}{2002}]{goto02}
Goto T., et al., 2002, PASJ, 54, 515 

\bibitem[\protect\citeauthoryear{Goto et al.}{2003}]{goto03}
Goto T.~et al., 2003, PASJ, 55, 757

\bibitem[\protect\citeauthoryear{Gunn \& Gott}{1972}]{gunn72}
Gunn J.E., Gott J.R., 1972, ApJ, 176, 1

\bibitem[\protect\citeauthoryear{Guzman et al.}{1997}]{guzman97}
Guzman R., Gallego J., Koo D.~C., Phillips A.~C., Lowenthal J.~D., Faber S.~M., Illingworth G.~D., Vogt N.~P., 1997, ApJ, 489, 559  

\bibitem[\protect\citeauthoryear{Hammer et al.}{2005}]{hammer05}
Hammer F., Flores H., Elbaz D., Zheng X.~Z., Liang Y.~C., Cesarsky C., 2005, A\&A, 430, 115

\bibitem[\protect\citeauthoryear{Hashimoto et al.}{1998}]{hashimoto98}
Hashimoto Y., Oemler A., Lin H., Tucker D.L., 1998, ApJ, 499, 589

\bibitem[\protect\citeauthoryear{Heavens et al.}{2004}]{heavens04}
Heavens A., Panter B., Jimenez R., Dunlop J., 2004, Nature, 428, 625

\bibitem[\protect\citeauthoryear{Hogg et al.}{2002}]{hogg02}
Hogg D.~W., Blanton M., Strateva I., et al., 2002, AJ, 124, 646

\bibitem[\protect\citeauthoryear{Hopkins et al.}{2003}]{hopkins03}
Hopkins A.~M., et al., 2003, ApJ, 599, 971 

\bibitem[\protect\citeauthoryear{Jimenez et al.}{2005}]{jimenez05}
Jimenez R., Panter B., Heavens A.~F., Verde L., 2005, MNRAS, 356, 495

\bibitem[\protect\citeauthoryear{Juneau et al.}{2005}]{juneau05}
Juneau S., et al., 2005, ApJ, 619, L135

\bibitem[\protect\citeauthoryear{Kaiser}{1984}]{kaiser84}
Kaiser N., 1984, ApJ, 284, L9  

\bibitem[\protect\citeauthoryear{Kauffmann et al.}{2003a}]{kauffmann03a}
Kauffmann G., et al., 2003a, MNRAS, 341, 33

\bibitem[\protect\citeauthoryear{Kauffmann et al.}{2004}]{kauffmann04}
Kauffmann G., White S.~D.~M., Heckman T.~M., M{\' e}nard B., Brinchmann J., Charlot S., Tremonti C., Brinkmann J., 2004, MNRAS, 353, 713

\bibitem[\protect\citeauthoryear{Kennicutt}{1998}]{kennicutt98}
Kennicutt R.~C., 1998, ARA\&A, 36, 189

\bibitem[\protect\citeauthoryear{Kewley, Jansen \& Geller}{Kewley et al.}{2005}]{kewley05} Kewley L.~J., Jansen R.~A., Geller M.~J., 2005, PASP, 117, 227

\bibitem[\protect\citeauthoryear{Kodama et al.}{2004}]{kodama04}
Kodama T., et al., 2004, MNRAS, 350, 1005

\bibitem[\protect\citeauthoryear{Koopmann \& Kenney}{2004}]{koopmann04}
Koopmann R.~A., Kenney J.~D.~P., 2004, ApJ, 613, 866
 
\bibitem[\protect\citeauthoryear{Larson, Tinsley \& Caldwell}{1980}]{LTC}
Larson R.B., Tinsley B.M., Caldwell C.N., 1980, ApJ, 237, 692

\bibitem[\protect\citeauthoryear{Lewis et al.}{2002}]{lewis02}
Lewis I., et al. (The 2dFGRS Team), 2002, MNRAS, 334, 673

\bibitem[\protect\citeauthoryear{Madau, Ferrara \& Rees}{2001}]{madau01}
Madau P., Ferrara A., Rees M.~J., 2001, ApJ, 555, 92

\bibitem[\protect\citeauthoryear{Martin \& Kennicutt}{2001}]{martin01}
Martin C.~L., Kennicutt R.~C., 2001, ApJ, 555, 301

\bibitem[\protect\citeauthoryear{Mateus \& Sodr\'e}{2004}]{mateus04}
Mateus A., Sodr\'e L., 2004, MNRAS, 349, 1251

\bibitem[\protect\citeauthoryear{SEAGal II}{}]{mateus06}
Mateus A., Sodr\'e L., Cid Fernandes R., Stasi\'nska G., Schoenell W., Gomes J.~M., 2006, MNRAS, 370, 721

\bibitem[\protect\citeauthoryear{McCarthy et al.}{2004}]{mccarthy04}
McCarthy P.~J., et al., 2004, ApJ, 614, L9 

\bibitem[\protect\citeauthoryear{Menci et al.}{2005}]{menci05}
Menci N., Fontana A., Giallongo E., Salimbeni S., 2005, ApJ, 632, 49

\bibitem[\protect\citeauthoryear{Miller et al.}{2003}]{miller03}
Miller C.~J., Nichol R.~C., G\'omez P.~L., Hopkins, A.~M., Bernardi M., 2003, ApJ, 597, 142

\bibitem[\protect\citeauthoryear{Miller et al.}{2005}]{miller05}
Miller C.~J., et al., 2005, AJ, 130, 968  

\bibitem[\protect\citeauthoryear{Murtagh \& Heck}{1987}]{murtagh87}
Murtagh F., Heck A., 1987, Multivariate data analysis. Astrophysics and
Space Science Library. Reidel, Dordrecht

\bibitem[\protect\citeauthoryear{Poggianti et al.}{1999}]{poggianti99}
Poggianti B.M., Smail I., Dressler A., Couch W.J., Barger A.J., Butcher H., Ellis R.S., Oemler A., 1999, ApJ, 518, 576

\bibitem[\protect\citeauthoryear{Poggianti et al.}{2004}]{poggianti04}
Poggianti B.~M., Bridges T.~J., Komiyama Y., Yagi M., Carter D., Mobasher B., Okamura S., Kashikawa N., 2004, ApJ, 601, 197  

\bibitem[\protect\citeauthoryear{Poggianti et al.}{2005}]{poggianti05}
Poggianti B., et al., 2005, ApJ, in press (astro-ph/0512391)

\bibitem[\protect\citeauthoryear{Ridley}{2003}]{ridley03}
Ridley M., 2003, Nature via nurture - Genes, experience and what makes us human, Harper
   Collins Publishers, New York

\bibitem[\protect\citeauthoryear{Rines et al.}{2005}]{rines05}
Rines K., Geller M.~J., Kurtz M.~J., Diaferio A., 2005, AJ, 130, 1482

\bibitem[\protect\citeauthoryear{Scannapieco, Ferrara \& Madau}{2002}]{scannapieco02} Scannapieco E., Ferrara A., Madau P., 2002, ApJ, 574, 590

\bibitem[\protect\citeauthoryear{Scannapieco, Silk, \& Bouwens}{2005}]{scannapieco05} Scannapieco E., Silk J., Bouwens R., 2005, ApJ, 635, L13

\bibitem[\protect\citeauthoryear{Schlegel, Finkbeiner, \& Davis}{Schlegel et al.}{1998}]{schlegel98}
Schlegel D.~J., Finkbeiner D.~P., Davis M., 1998, ApJ, 500, 525

\bibitem[\protect\citeauthoryear{Sodre \& Cuevas}{1997}]{sodre97}
Sodre L., Cuevas H., 1997, MNRAS, 287, 137

\bibitem[\protect\citeauthoryear{Solanes et al.}{1996}]{solanes96}
Solanes J.M., Giovanelli R., Haynes M.P., 1996, ApJ, 461, 609

\bibitem[\protect\citeauthoryear{Steidel et al.}{2005}]{steidel05}
Steidel C.~C., Adelberger K.~L., Shapley A.~E., Erb D.~K., Reddy N.~A., Pettini M., 2005, ApJ, 626, 44

\bibitem[\protect\citeauthoryear{Stoughton et al.}{2002}]{stoughton02}
Stoughton C., et al., 2002, AJ, 123, 485

\bibitem[\protect\citeauthoryear{Strateva et al.}{2001}]{strateva01}
Strateva I., et al., 2001, AJ, 122, 1861 

\bibitem[\protect\citeauthoryear{Strauss et al.}{2002}]{strauss02} Strauss M.~A., et al., 2002, AJ, 124, 1810

\bibitem[\protect\citeauthoryear{Tanaka et al.}{2004}]{tanaka04}
Tanaka M., Goto T., Okamura S., Shimasaku K., Brinkmann J., 2004, AJ, 128, 2677

\bibitem[\protect\citeauthoryear{Tanaka et al.}{2005}]{tanaka05}
Tanaka M., Kodama T., Arimoto N., Okamura S., Umetsu K., Shimasaku K., Tanaka I., Yamada T., 2005, MNRAS, 362, 268

\bibitem[\protect\citeauthoryear{Thomas et al.}{2005}]{thomas05}
Thomas D., Maraston C., Bender R., de Oliveira C.~M., 2005, ApJ, 621, 673  

\bibitem[\protect\citeauthoryear{Tran et al.}{2003}]{tran03}
Tran K.-V.~H., Franx M., Illingworth G., Kelson D.~D., van Dokkum P., 2003, ApJ, 599, 865  

\bibitem[\protect\citeauthoryear{Tully}{2005}]{tully05}
Tully R.~B., 2005, ApJ, 618, 214

\bibitem[\protect\citeauthoryear{Vollmer et al.}{2001}]{vollmer01}
Vollmer B., Cayatte V., Balkowski C., Duschl W.J., 2001, ApJ, 561, 708

\bibitem[\protect\citeauthoryear{Weinberg et al.}{2004}]{weinberg04}
Weinberg D.~H., Dav{\'e} R., Katz N., Hernquist L., 2004, ApJ, 601,~1

\bibitem[\protect\citeauthoryear{Whitmore, Gilmore \& Jones}{1993}]{whitmore93} 
Whitmore B.C., Gilmore D.M., Jones C., 1993, ApJ, 407, 489.

\end{thebibliography}
\end{document}